\documentclass[preprint,aps,pre,floatfix,linenumbers]{revtex4-1}
\usepackage{graphicx}
\usepackage{epstopdf}
\usepackage{color}
\usepackage[T1]{fontenc}
\usepackage{longtable}

\newcommand{\Rm}{\mbox{Rm}}
\newcommand{\Pm}{\mbox{Pm}}
\newcommand{\Ha}{\mbox{Ha}}
\renewcommand{\Re}{\mbox{Re}}

\renewcommand{\deg}{^{\circ}}
\usepackage{mathcomp}

\begin{document}
\title{Zonal shear and super-rotation in a magnetized spherical Couette flow experiment}

\author{D. Brito}
\affiliation{Laboratoire des Fluides Complexes et leurs R\'eservoirs, Universit\'e de Pau et des Pays de l'Adour, CNRS, BP 1155, 64013 Pau Cedex France}
\altaffiliation{formerly member of the geodynamo team (Institut des Sciences de la Terre)}
\email{daniel.brito@univ-pau.fr}
\author{T. Alboussi\`ere}
\altaffiliation{now at : Laboratoire de Sciences de la Terre, Université de Lyon, ENS de Lyon, CNRS, Lyon, France}
\author{P. Cardin}
 \author{N. Gagni\`ere}
 \author{ D. Jault}
 \author{P. La Rizza}
 \author{J.-P. Masson}
 \author{H.-C. Nataf} 
 \author{D. Schmitt}
\affiliation{Institut des Sciences de la Terre,
CNRS, Observatoire de Grenoble, Universit\'e Joseph-Fourier, Maison
des G\'eosciences, BP 53, 38041 Grenoble Cedex 9, France}

\date{\today}

\begin{abstract}
We present measurements performed in a spherical shell
filled with liquid sodium, where a 74 mm-radius inner sphere is
rotated while a 210 mm-radius outer sphere is at rest. The inner
sphere holds a dipolar magnetic field and acts as a magnetic propeller
when rotated. In this
experimental set-up called $DTS$, direct measurements of the
velocity are performed by ultrasonic Doppler velocimetry.
Differences in electric potential and the induced magnetic field
 are also measured to characterize the magnetohydrodynamic
flow. Rotation frequencies of the inner sphere
are varied between -30 Hz and +30 Hz, the magnetic Reynolds number based on measured sodium velocities
and on the shell radius
reaching to about 33. We have investigated the mean
axisymmetric part of the flow, which consists of differential rotation. Strong super-rotation of the fluid with respect to 
the rotating inner sphere is directly measured. It is found that the organization of the mean flow
does not change much throughout the entire range of
parameters covered by our experiment.
The direct measurements of zonal velocity give a nice illustration of Ferraro's law of isorotation
in the vicinity of the inner sphere where magnetic forces dominate inertial ones. The transition from a Ferraro regime in the interior to a geostrophic regime, where inertial forces predominate, in the outer regions has been well documented.
It takes place where the local Elsasser number is about 1.
A quantitative agreement with non-linear numerical simulations is obtained when keeping the same Elsasser number.
The experiments also reveal a region that violates Ferraro's law just above the inner sphere. 
\end{abstract}

\maketitle

\section{Introduction}

The Earth's fluid core below the solid mantle consists of a 3480 km-radius
spherical cavity filled with a liquid iron alloy. A 1220 km-radius
solid inner core sits in its center. It has been accepted since the 1940's \cite{elsasser46a,elsasser46b} that
the flows stirring the electrically conducting liquid iron in the outer core produce the Earth's magnetic field by dynamo action.
The fluid motion is thought to originate from the cooling of the Earth's core, which results both
in crystallization of the inner core and in convection in the liquid outer core \cite{verhoogen80}.

The last decade has seen enormous progress in the numerical
computation  of the geodynamo problem after the first simulation of
a dynamo powered by convection \cite{glatzmaier95,christensen2006,takahashi2008,sakuraba2009}. It is however still unclear
why many  characteristics of the Earth's magnetic field are so-well
retrieved with simulations \cite{dormy00} since the latter are
performed with values of  important dimensionless parameters that differ much from the appropriate values for the
Earth's core. The main numerical difficulty is the simultaneous computation of
the velocity, the magnetic and the temperature fields
with realistic diffusivities, respectively the fluid viscosity, the
magnetic and the thermal diffusivities. Those
differ indeed by six orders of magnitude in the
outer core \cite{poirier00}; such a wide range is at present out of reach
numerically, the simulations being performed at best with two orders
of magnitude difference between the values of the diffusivities. An experimental
approach of the geodynamo is, in that respect, promising since
the fluid metals used in experiments have physical
properties, specifically diffusivities, very close to the properties of the liquid iron alloy in
the Earth's outer core. Moreover, experiments and
simulations are complementary since they span different
ranges of dimensionless parameters.

Magnetohydrodynamics experiments devoted to the dynamo study have
started some 50 years ago (see the chapter authored by Cardin and Brito in \cite{dormy07}
for a review). To possibly induce magnetic fields, the
working fluid must be liquid sodium in such experiments. Sodium is indeed the
fluid that best conducts electricity in laboratory conditions. A breakthrough
in these dynamo experiments occurred at the end of 1999 when
amplification and  saturation of an imposed magnetic field were
 measured for the first time in two experiments, in Riga \cite{gailitis01} and in
Karlsruhe \cite{stieglitz01}. The commun property of those set-ups was
to have the sodium motion very much constrained spatially, in order to closely
follow  fluid flows well known analytically to lead to a kinematic dynamo,
respectively the Ponomarenko flow \cite{ponomarenko73} and the
G.O. Roberts flow \cite{roberts72}. More recently, the first
experimental dynamo in a fully turbulent flow was obtained  in a
configuration where two crenelated ferromagnetic rotating discs drive a von K\`arm\`an swirling flow
in a cylinder \cite{berhanu07}.
Earth's like magnetic field reversals were also obtained in this
experimental dynamo \cite{monchaux07}. Other similar experiments have been run where  sodium flows are driven by propellers in a
spherical geometry \cite{sisan04,nornberg06}.
In order to emphasize the specificity of the experimental study
presented in the present paper, it is worth mentioning two common features of
the previously mentioned sodium experiments: the forcing of the sodium motion is always purely
mechanical and the magnetic field is \emph{weak} in
the sense that Lorentz forces are small compared to the non-linear
velocity terms in the equation of motion \cite{petrelis07}.

The experiment called $DTS$ for "Derviche Tourneur Sodium" has
been designed to investigate a supposedly relevant regime for the
Earth's core, the magnetostrophic regime \cite{taylor63,cardin02,jault08} where the
ratio of Coriolis to Lorentz forces is of the order one. The
container made of weakly conducting stainless steel is spherical and can rotate about a vertical axis.
An inner sphere consisting of a copper envelope enclosing
permanent magnets is placed at the center of the outer sphere; the
force free magnetic field produced by those magnets enables to explore dynamical regimes where Coriolis and Lorentz
forces are comparable. The sodium motion in the spherical
gap is driven by the differential rotation between the inner sphere
and the outer sphere, unlike in the Earth's core where the iron
motion is predominantly driven by convection
\cite{busse70} and maybe minorly  by differential rotation of the inner core
\cite{song96}.

The $DTS$  experiment has not been designed to run in a dynamo
regime. It has instead been conceived as a 
small prototype  of a possible future large sodium spherical dynamo experiment
which would benefit from its results. Note that meanwhile Daniel
Lathrop and collaborators have built a 3m-diameter
sodium spherical experiment with an
inner sphere differentially rotating with respect to the outer
sphere, like in $DTS$. Schaeffer, Cardin and Guervilly \cite{schaeffer06,guervilly10} have shown numerically that a
dynamo could occur in a spherical Couette flow at large Rm in a low
magnetic Prandtl number fluid such
as sodium (Pm$= \nu / \lambda$, see TABLE~\ref{properties}).

Numerical simulations in a $DTS$-type configuration
\cite{Hollerbach94,dormy98,hollerbach07}
of Couette spherical flows with an imposed magnetic field all show azimuthal flows stabilized by
magnetic and rotation forces.
Using electric potential measurements along a meridian of the outer sphere boundary,
we concluded in our first report of  $DTS$ experimental results \cite{nataf06} that the
amplitude of the azimuthal flow may exceed
the velocity of sodium in solid body rotation with the inner sphere, as 
predicted theoretically in the linear regime \cite{dormy02}.

The $DTS$ experiment offers a tool to investigate
non uniform rotation of an electrically conducting fluid in the presence of rotation and magnetic forces.
The differential rotation of a body permeated by a strong magnetic field and the waves driven by the
non uniform rotation have received considerable attention
since the work of Ferraro \cite{ferraro37,spruit99}. Indeed, the absence of solid envelopes makes non uniform rotation possible in stars,
where it plays an important role in the mixing of chemical elements \cite{charbonnel2005influence}, 
in contrast with the case of  planetary fluid cores. Ferraro found that the angular rotation in an electrically conducting
body permeated by a steady magnetic field symmetric about the axis of rotation tends to be constant
along magnetic lines of force. 
MacGregor and Charbonneau \cite{macgregor1999angular} illustrated this result and showed, in a weakly rotating case, that Ferraro's theorem holds for $\Ha \gg 1$ ($\Ha$, the Hartmann number, measures the magnetic strength, see TABLE~\ref{sansdim}).
An intense magnetic field, probably of primordial origin, is the key actor in the transfer of angular momentum from the solar radiative interior to the convection zone \cite{mestel1987,gough2010}. 
Finally, in a geophysical context, Aubert recently found, investigating zonal flows in spherical shell dynamos,  that Ferraro's law of isorotation gives a
good description of the geometry of the zonal flows of thermal origin \cite{aubert2005steady}.

In the second study of the $DTS$ experiment \cite{nataf07}, we investigated azimuthal
flows when both the inner boundary and the outer boundary are rotating but at different speeds, using Doppler velocimetry and electric potential measurements. Specifically, we discussed the 
transition between the outer geostrophic region and the inner region where magnetic forces dominate. 
Extending the asymptotic model of Kleeorin et al. \cite{kleeorin97}, we could explain the shape of the measured azimuthal velocity profiles.
We had to use a specific electric
potential difference as a proxy of the differential rotation between the two spheres as, unfortunately, the electrical
coupling between the liquid sodium and the copper casing of the interior magnets was apparently both imperfect and unreliable.
Finally, we reported in on our third article \cite{schmitt07} about the $DTS$ experiment the presence of azimuthally traveling hydromagnetic waves
that we inferred mainly  from electric potential measurements along parallels.

We investigate here again the main flows when the outer sphere is at rest. Our new study benefits from a comparison with our earlier work \cite{nataf07} for a rotating outer sphere. There is no need any more to use an indirect measure of the global rotation of the fluid as  the electrical coupling between liquid sodium and copper has become unimpaired.
Furthermore, the $DTS$ experiment has been equipped with a host of new  measurement tools. The flow amplitude is measured along 7 different beams using Doppler velocimetry. Assuming axisymmetry, we have thus been able to map the azimuthal flow in most of the fluid. It turns out that the electric potential differences evolve monotonically with the inner core rotation but cannot be interpreted directly as a measure of the velocity below the outer viscous boundary layer. We have also entered a probe inside the cavity
to measure the induced magnetic field in the interior. 
The dense measurements in the $DTS$ experiment give a nice illustration of the Ferraro law of isoration \cite{ferraro37}
in the inner region where magnetic forces dominate. In the outer region, we retrieve axially invariant azimuthal flow as the Proudman-Taylor theorem holds there, even though the outer sphere is at rest. The variation of the geostrophic velocity with the distance to the axis differs nevertheless from the case of a rotating outer sphere as recirculation in the outer Ekman layer plays an important role in the latter case.

The organisation of the paper is as follows. In section
\ref{dts}, we describe the experimental set-up and the 
techniques that we use to measure the magnetic, electric and velocity fields; we illustrate them with a discussion of a typical experimental run. In section \ref{equations}, we present the
governing equations and the relevant dimensionless numbers of the
experiment.  We devote one section of the article to the observation of differential rotation and another one to
the meridional circulation. Then,
the experimental measurements are 
compared to numerical simulations of $DTS$. We summarize
and discuss the results of our study in section \ref{conclusion}.

%\clearpage

\clearpage
\section{\label{dts}The $DTS$ Experiment}
\subsection{\label{set_up}The experimental set-up}

\begin{figure}[h]
\centerline{{\includegraphics[width=0.53\linewidth]{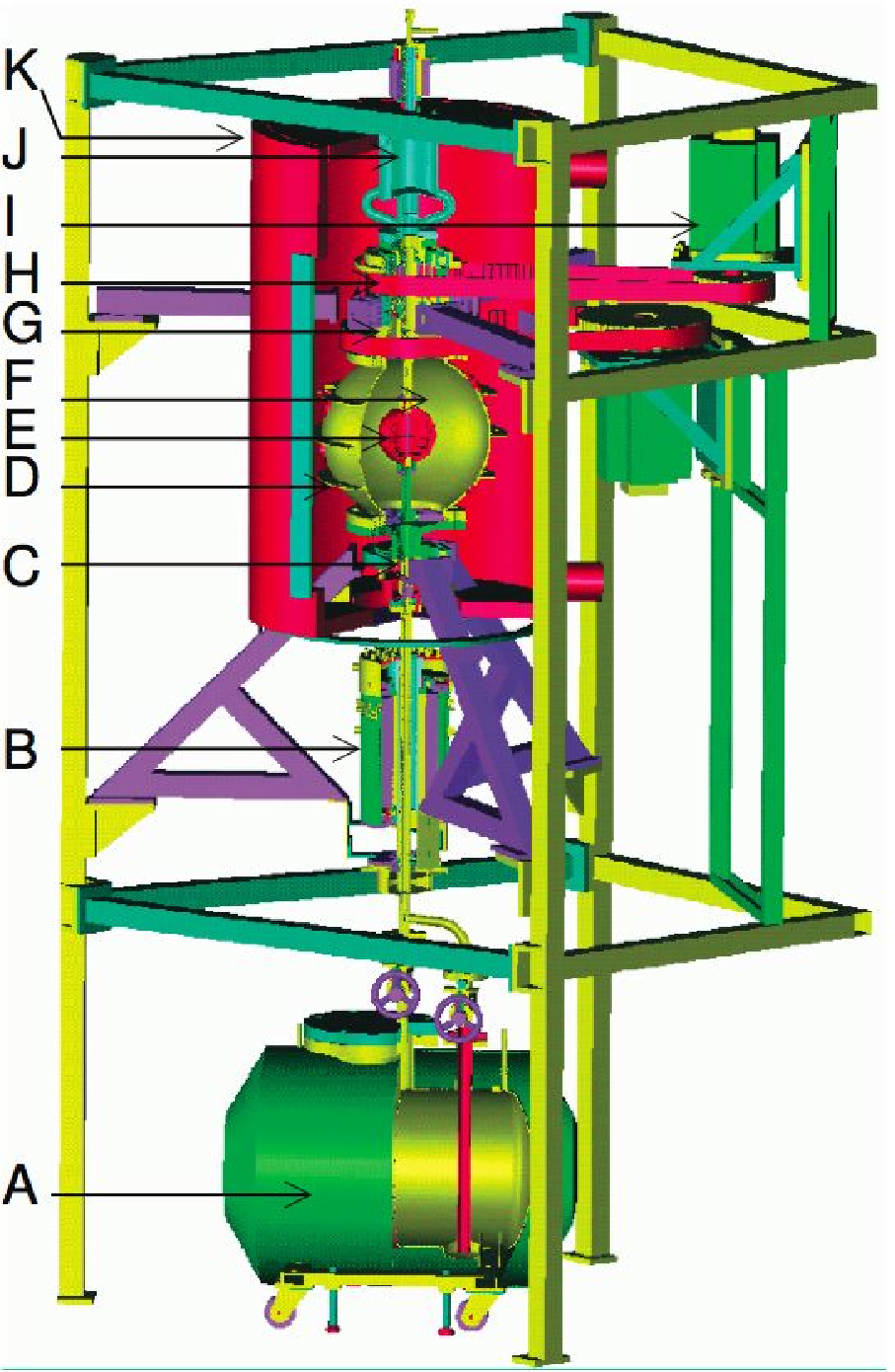}}
{\includegraphics[width=0.47\linewidth]{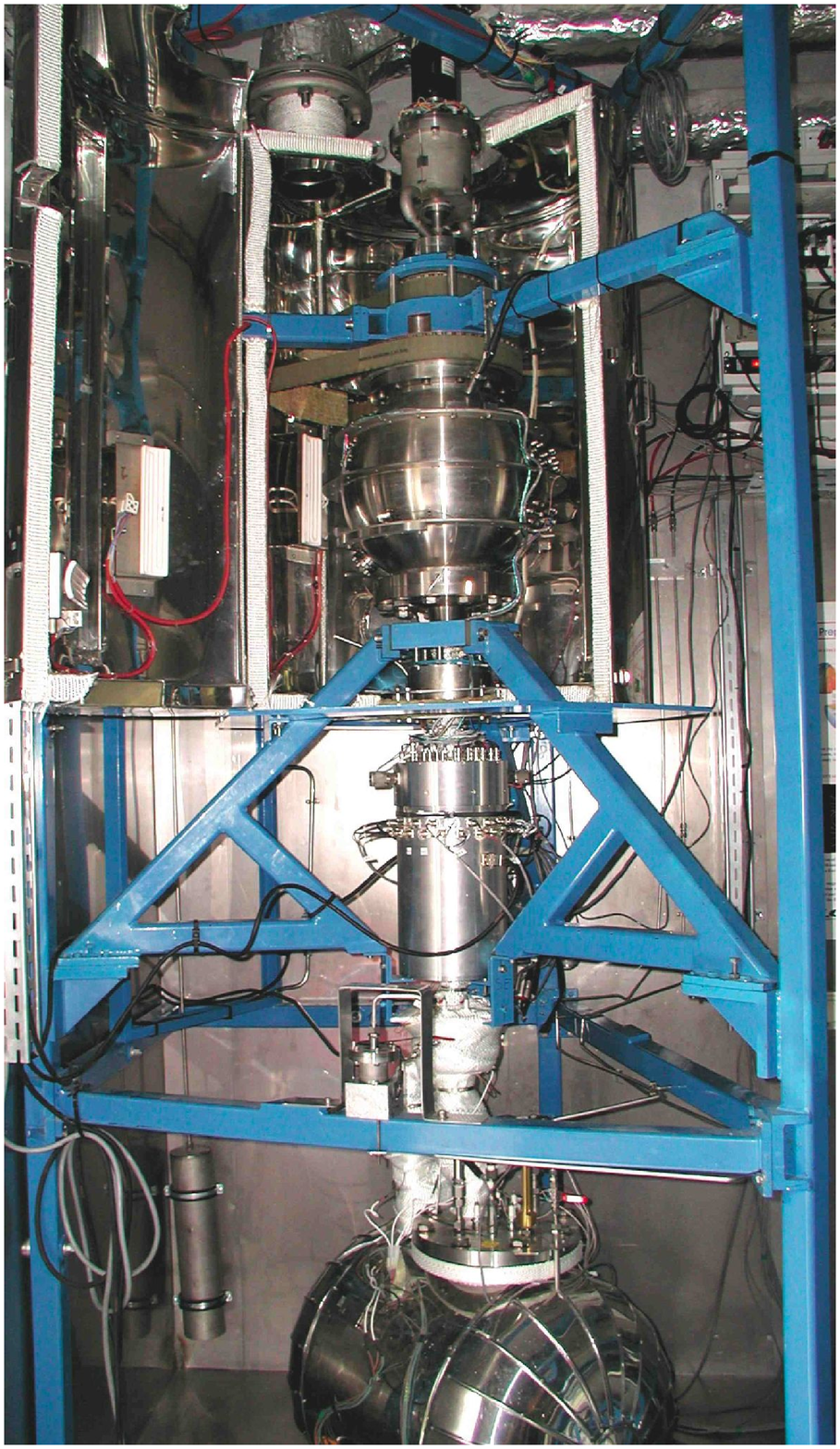}}}
\caption{(Color online) Diagram and picture of the experimental set--up. A: moveable sodium
reservoir, B: shielded electric slip-ring, C:
electromagnetic valve, D: outer sphere, E: magnetized rotating inner
sphere, F: spherical shell containing liquid sodium, G: magnetic
coupling entraining the inner sphere shaft, H: crenelated belt, I:  brushless electric
motor driving the inner sphere, J: expansion tank for sodium, K:
thermostated chamber.  The total height of the set-up is 3.9 m. \label{Experiment}}
\end{figure}

The $DTS$ experimental set-up \cite{nataf06,nataf07,schmitt07}  
is shown in FIG.~\ref{Experiment}. It has been installed in a small building
purpose-designed for sodium experiments.

\begin{figure}[hp]
\centerline{\includegraphics[width=0.263\linewidth]{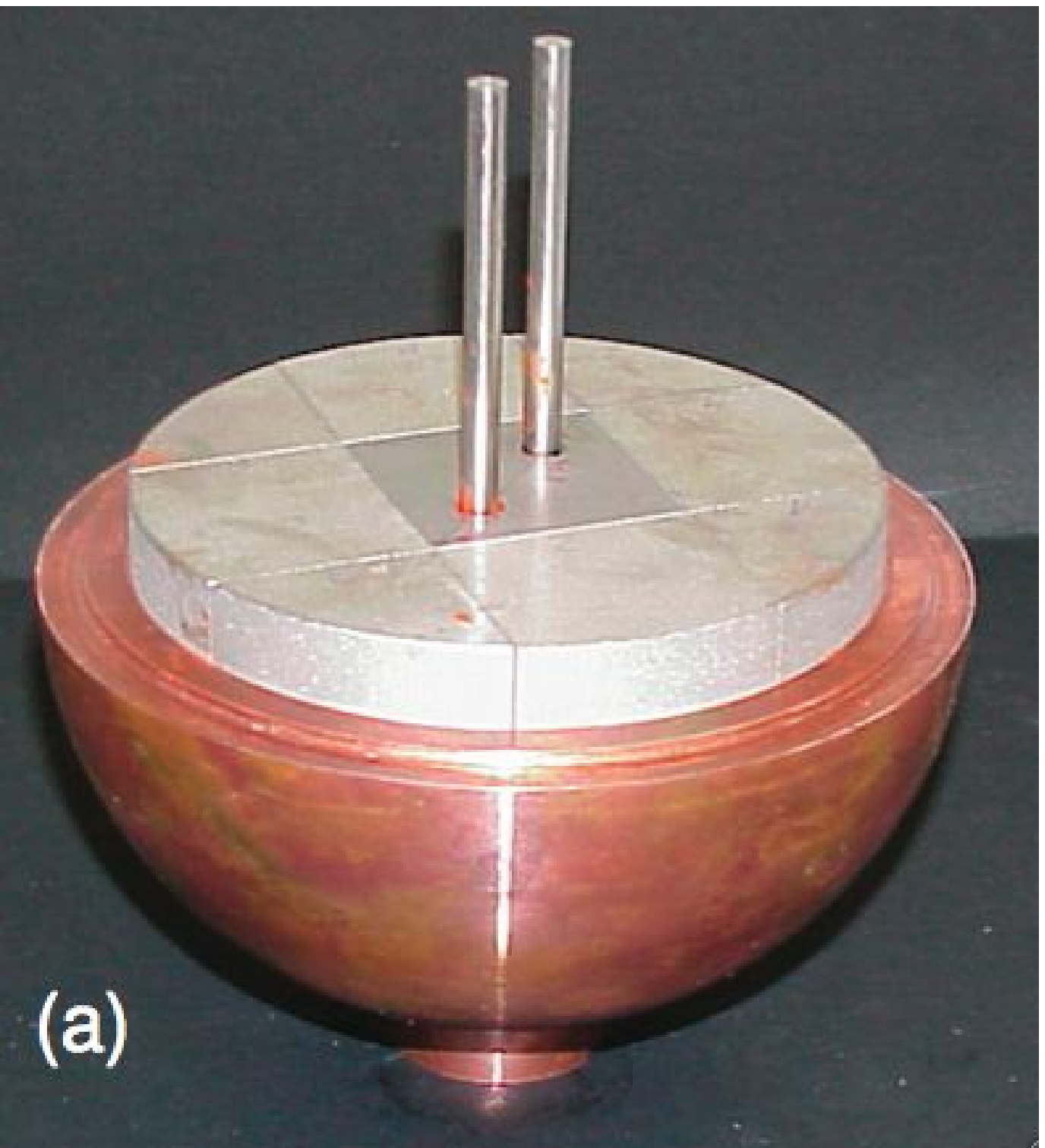}  \includegraphics[width=0.73 \linewidth]{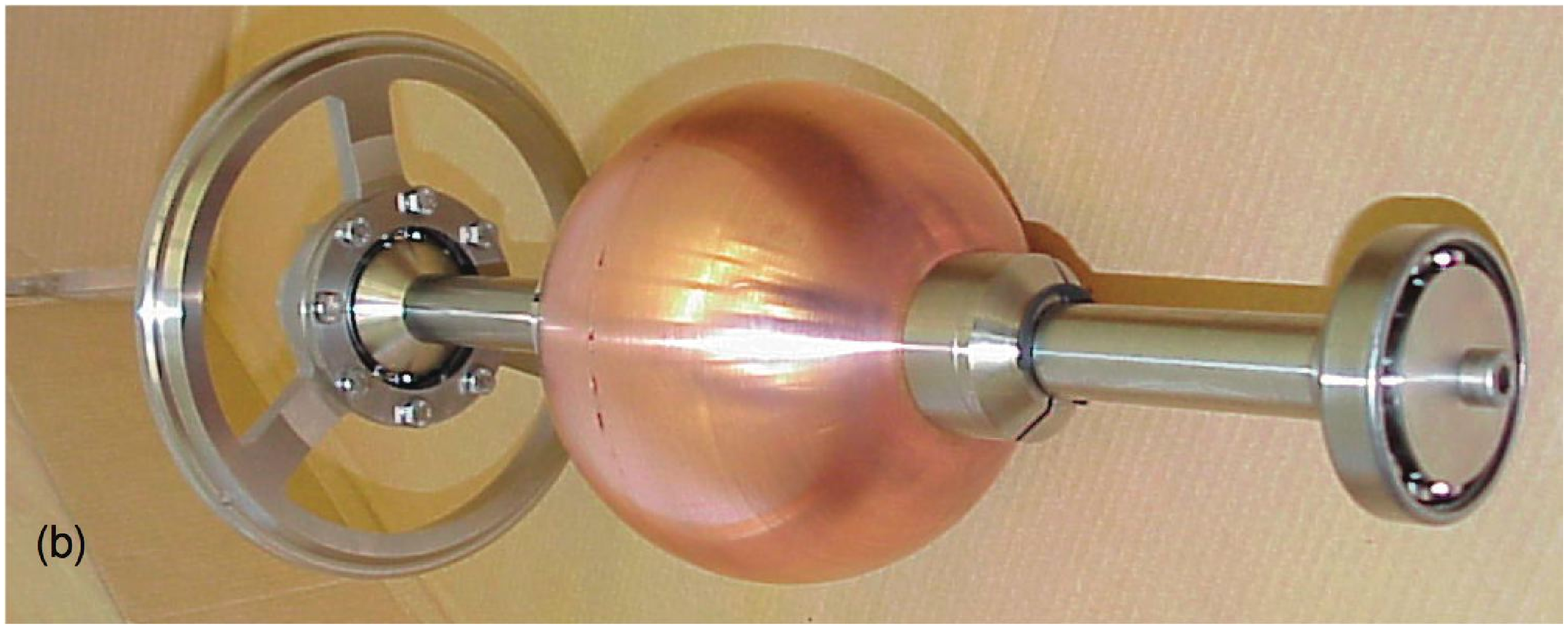}}
\caption{(Color online) (a) Picture of one hemisphere of the inner sphere.
Different pieces of magnets in gray are assembled in the bulk of the
inner sphere. 
(b) View from the side of the inner sphere and its rotating shaft.
Note that the wheels at the top and bottom (only one is shown in the picture) of the rotating shaft are attached to the outer
sphere.}
 \label{graine}
\end{figure}

As shown in FIG.~\ref{Experiment}, liquid sodium is contained in a
spherical shell between an outer sphere and an inner sphere.
The radius of the outer sphere is $a$ = 210 mm and
that of the inner sphere $b$ = 74 mm. The outer sphere is made of
stainless steel and is 5 mm thick. The copper inner sphere (FIG.~\ref{graine} and FIG.~\ref{trajectoires}) contains magnetized Rare-Earth cobalt
bricks assembled such that the resulting permanent magnetic
field is very close to an axial dipole of moment intensity 700 Am$^2$, with its axis of
symmetry aligned with the axis of rotation.
The magnetic field points upward along the rotation axis
and its magnitude ranges from 345 mT at the poles of the inner
sphere down to 8 mT at the equator of the outer sphere.

Sodium is kept most of the time in the reservoir at the bottom of
the set-up. When needed to run an experiment,
sodium is melted and pushed up from that reservoir into the spherical shell by
imposing an overpressure of Argon in the reservoir. When liquid
sodium reaches the expansion tank at the top of the spherical shell,
an electromagnetic valve located just below the sphere (see
FIG.~\ref{Experiment}) is locked such that sodium is kept in the
upper part during experiments. In case of emergency, the valve is
opened and sodium pours directly into the reservoir.

The central part of the experiment is air-conditioned in a chamber
maintained at  around $130 \tccentigrade$ during experiments: four 1 kW infrared radiants disposed
around the outer sphere heat the chamber,  whereas cold
air pumped from outside cools the set-up when necessary. Liquid
sodium is therefore usually kept some 30$\tccentigrade$ above its melting temperature during experiments.
Some physical properties of sodium
relevant to our study are listed in TABLE~\ref{properties}. The
whole volume containing sodium, from the reservoir tank up to the
expansion tank is kept under Argon pressure at all times in order to
limit oxidization of sodium.

\begin{table}
\caption{\label{properties}Physical properties of pure liquid sodium at $130
\tccentigrade$ (Documents from CEA, Commissariat \`a l'Energie Atomique et aux \'energies alternatives). *The sound velocity in sodium has been precisely measured in the present study using the UDV apparatus.}
\begin{ruledtabular}
\begin{tabular}{cllc}
 $\rho$  & density  & 9.3 $10^{2}$ kg m$^{-3}$  \\
 $\sigma$& electric conductivity& $9\;10^{6} \; \Omega^{-1}$m$^{-1}$  \\
 $\nu$ & kinematic viscosity & $6.5 \;10^{-7}$ m$^2$s$^{-1}$  \\
 $\eta$ & magnetic diffusivity &8.7 $10^{-2}$ m$^2$s$^{-1}$  \\
 $c$ & sound velocity* & 2.45 $10^{3}$ m s$^{-1}$   \\
\end{tabular} 
\end{ruledtabular}
\end{table}

The rotation of the inner sphere, between $ f=-30$ Hz and
$ f=30$ Hz, is driven by a crenelated belt attached to a 11 kW
brushless motor (SGMH-1ADCA61 from Yaskawa Electric Corporation,
Tokyo, Japan). The belt entrains a
home-made magnetic coupler located around the inner sphere shaft as
seen in FIG.~\ref{Experiment}. The coupler is composed of an
array of  magnets located outside the sodium
container, another array of magnets inside the container being
immersed in liquid sodium. The inner magnets  are anchored to the
rotating shaft of the inner sphere such that when the belt is
rotated outside, the inner sphere is rotated as well. Such a
coupler has the advantage of not requiring any rotating seal in
liquid sodium. Torque values up to about 70 N$\cdot$m  have been
efficiently transmitted through this coupler in the experiment.

\subsection{Measurements}

\subsubsection{Ultrasonic Doppler
velocimetry} \label{partUDV} We use UDV ultrasonic Doppler velocimetry
\cite{Takeda87} in order to measure liquid sodium velocities in
the spherical shell. This non intrusive technique has been intensively used in our
group for the last decade, in particular in rotating
experiments performed either in water or in liquid metals \cite{brito01,noir01,aubert01,gillet07}. The  technique 
consists in the emission from a piezoelectric transducer of a succession of bursts of ultrasonic
waves that propagate in the fluid. When the wave encounters a particle with a
different acoustic impedance,
part of the ultrasonic wave is backscattered towards the transducer.
The time elapsed between the emitted and the reflected waves and the change in that time respectively give the
position of the particle with respect to the transducer and the fluid velocity along the beam direction. Data
processing is internal to the DOP2000 apparatus
(\texttt{http://www.signal-processing.com}, Signal Processing
company, Lausanne, Switzerland).

\begin{figure}[h]
\centerline{\includegraphics[width=0.60\linewidth]{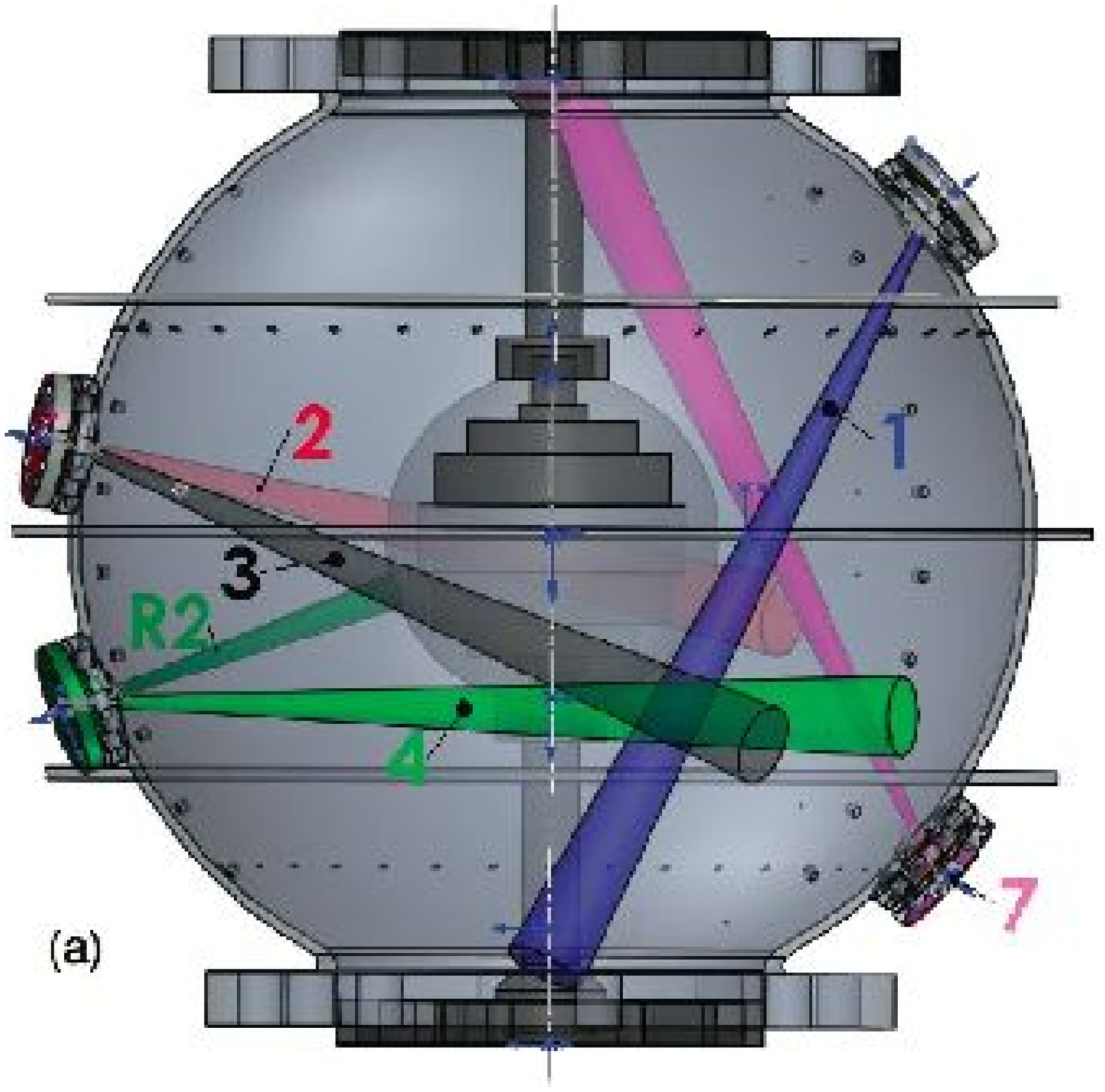}\includegraphics[width=0.31\linewidth]{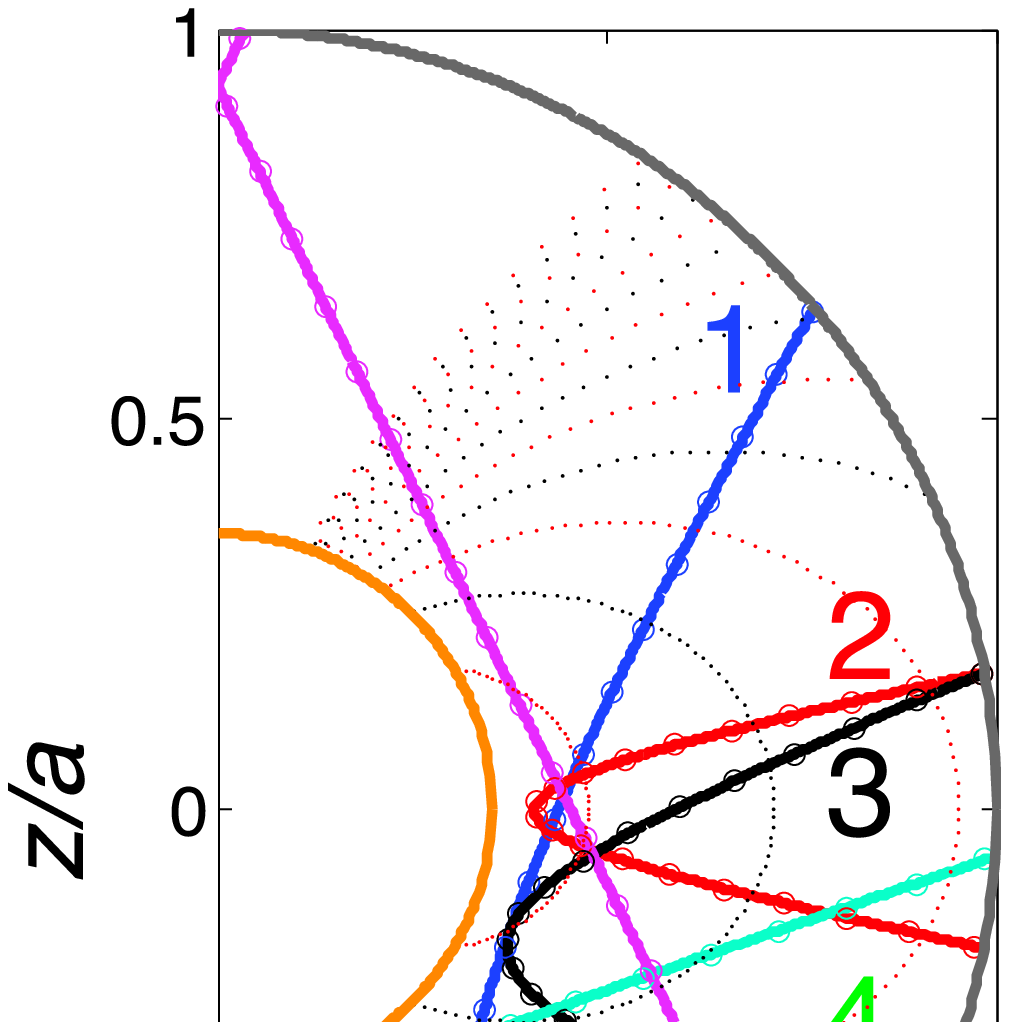}}
\caption{(Color online) (a) 3D perspective view of the outer sphere and its interior.  Caps at various latitudes  hold ultrasonic velocity probes to perform UDV. The divergent ultrasonic beams emitted from each cap are shown in perspective with different colors (and numbers for the grayscale version). The five superimposed horizontal slices
of magnets are assembled in the heart of the inner sphere. Differences in electric potential are measured between
points from latitude  +45$^{\circ}$ to latitude -45$^{\circ}$, with steps of
10$^{\circ}$ (holes along a meridian at the right of the Figure). (b) Meridional view of the normalized coordinates $(s/a,z/a)$ covered by the ultrasonic trajectories numbered from 1 to 7. Some of the corresponding rays are plotted in (a) with the same color code (same numbers). The distance $d$ from the outer sphere along the ultrasonic beam is marked by small
dots drawn every 20 mm. The dotted lines are field lines of the imposed dipolar magnetic field. }
\label{trajectoires}\end{figure}

The ultrasonic probes are held in circular stainless steel caps attached to the outer sphere, as shown in FIG.~\ref{trajectoires}(a). There
are six locations with interchangeable caps on the outer sphere such
that fluid velocities can be measured  from any of these different positions.
The thickness of the stainless steel wall between the
probes and liquid sodium has been very precisely machined
to 1.4 mm in order to insure the best transmission of energy from
the probe to the fluid \cite{eckert02}. Small sodium
oxides and/or gas bubbles are present and backscatter ultrasonic waves as in gallium
experiments \cite{gillet07}.
We keep the surface of the caps in contact with
sodium as smooth and clean as possible to perform UDV measurements.

We use high temperature 4 MHz ultrasonic transducers (TR0405AH from
Signal Processing) 10 mm long and 8 or 12 mm in diameter
(piezoelectric diameter 5 or 7 mm). The measurements shown throughout
the paper were performed with pulse repetition frequency ({\it prf})
varying from 3 kHz to 12 kHz and with a number of {\it prf} per profile
varying from 8 to 128. A present limitation of this
UDV technique is that the maximum measurable velocity obeys the following function
${{u}}_{\mbox{\tiny{max}}}= c^2 / 4 f_e P_{\mbox{\tiny{max}}} $
where  $c$ is the ultrasonic velocity of the medium, $f_e$ is the
emitting frequency, and $P_{\mbox{\tiny{max}}}$ is the maximum
measurable depth along the velocity profile. Applying this relationship to the
parameters used in $DTS$, $P_{\mbox{\tiny{max}}} \simeq 200$ mm
(approximative length of the first half of the beam in Figure
\ref{trajectoires}) and $f_e=4$ Mhz, the maximum measurable velocity is of
the order 2.2 m/s.
In particular cases, it is possible to overcome this limitation by
using aliased profiles of velocity \cite{brito01} as shown later in the paper.
The spatial resolution of the velocity profiles is about 1 mm, and the velocity resolution is about 0.5$\%$, or better for the aliased profiles.

We have measured  both the radial and oblique components of velocity
in the bulk of the spherical shell. The radial measurements were performed from the latitudes +10$^{\circ}$,
-20$^{\circ}$ and -40$^{\circ}$. The oblique measurements were performed from different locations and in different planes, along rays that all deviate from the radial direction by the same angle (24$^{\circ}$). Thus, they all have the same length in the fluid cavity. At the point of closest approach, the rays are 11 mm away from the inner sphere. 
The seven oblique beams used in $DTS$ are sketched in FIG.~\ref{trajectoires}(b). The way to retrieve the meridional and azimuthal components of the velocity field along the ultrasonic beam is detailed in the Appendix.

We use UDV measurements to confirm the strong magnetic coupling between the inner rotating sphere and sodium. In a smaller version of $DTS$ performed in water, maximum angular velocities (normalized by that of the inner sphere) of the order 0.16 are obtained for a hydrodynamic Reynolds number of $10^{5}$ in the vicinity of the equatorial plane, close to the rotating inner sphere \cite{guervilly10}. For similar $\Re$ in $DTS$, sodium is in super-rotation close to the inner rotating sphere and maximum measured velocities are instead around 1.2 (see FIG.~\ref{isolines}(b) for example). 

\subsubsection{\label{doigt_de_gant}Magnetic field inside the sphere}
The measurement technique described so far does not requires probes that protrude inside the sphere. In order to measure the magnetic field inside the sphere, in the liquid, we have installed magnetometers inside a sleeve, which enters deep into the liquid. The external dimensions of the sleeve are 114 mm (length inside the sphere) and 16 mm (diameter). It contains a board equipped with high-temperature Hall magnetometers (model A1384LUA-T of Allegro Microsystems Inc). We measure the radial component of the magnetic field at radii (normalized by $a$ the inner radius of the outer sphere) 0.93 and 0.74. The orthoradial component is measured at 0.97 and 0.78, and the azimuthal component at 0.99, 0.89, 0.79, 0.69, 0.60 and 0.50.
The sleeve is mounted in place of a removable port (at a latitude of either $40\deg$, $10\deg$ or $-20\deg$). A top view of the sleeve is shown in FIG.~\ref{inducedB}. 
The measured voltage is sampled at
2000 samples/second with a 16-bit 250 kHz PXI-6229 National
Instruments acquisition card.
The precision of the measurements (estimated from actual measurements when $f=0$) is about 140 $\mu$T, and corresponds to about 20 unit bits of the A/D converter. Magnetic fields up to 60 mT have been measured.

\subsubsection{\label{electrodes}Differences in electric potentials on the outer
sphere} 

Differences in electric potentials are measured along several
meridians and along one parallel of the outer sphere
\cite{nataf06,nataf07,schmitt07}. In the present
study, we are interested in the measurements performed along 
meridians since they are  linked to the azimuthal
flow velocity $u_{\varphi}$ 
(we denote $(r,\theta,\varphi)$ the spherical coordinates). 
The
measurements are performed between successive electrodes located from
-45$\deg$ to +45$\deg$ in latitude, with electrodes 10$\deg$ apart
as sketched in FIG.~\ref{trajectoires}(a).
We note $\Delta V_{40} = V_{45}-V_{35}$ the difference between the electric potential at latitudes $45\deg$ and $35\deg$.
Electric potentials are
measured by electrodes soldered to brass bolts 3 mm long, those
being screwed into 1 mm-diameter, 4 mm-deep blind holes drilled in
the stainless steel wall of the outer sphere. The measured voltage
is filtered by an
RC anti-aliasing 215 Hz low-pass filter and then sampled at
1000 samples/second with a 16-bit 250 kHz PXI-6229 National
Instruments acquisition card.
The precision of the measurements (estimated from actual measurements at $f=0$) is about 80 $\mu$V,
and corresponds to about 10 unit bits of the A/D converter.
Electric potential differences up to 7 mV have been measured.

Denoting $\mathbf{E}$ the electric field, we introduce the electric potential $V$
through $\mathbf{E}=-\mathbf{\nabla} V$, which is valid in a steady state.
Then, the electric potential  measurements are analysed using Ohm's law for
a moving conductor, 
$\mathbf{j} = \sigma \left ( \mathbf{u} \times \mathbf{B} + \mathbf{E} \right )$
where $\sigma$ is the electric conductivity,
$\mathbf{j}$ the electric current density vector,  $\mathbf{u}$
the velocity field  and  $\mathbf{B}$ the magnetic field.
If the meridional
electric currents $j_{\theta}$ are small compared to $ \sigma u_{\varphi}B_r$  in the fluid interior and away from the equatorial plane where $B_r=0$, and if the viscous boundary layer adjacent to the
outer sphere is thin, which ensures the continuity of $E_\theta$ through the layer, then the measured
differences in electric potential depend on the product of the
local radial magnetic field $B_r$ by $u_{\varphi}$, the azimuthal fluid velocity:
\begin{equation}
\frac{\Delta V}{a \Delta \theta} = u_{\varphi}B_r \; ,
\label{Ohm2}
\end{equation}
where $\Delta \theta=10^\circ$ is the angle between two electrodes.
However, we shall question below the assumption on the smallness of $j_{\theta}$, referred to as the frozen flux
hypothesis.

\subsubsection{Velocity and torque measured from the motor driving the inner sphere}

The electronic drive of the motor entraining the inner sphere delivers an analog signal for its angular velocity and its torque. We checked and improved the velocity measurement by calibrating it using a rotation counter, which consists of a small magnet glued on the entrainment pellet and passing once per turn in front of a magnetometer. The torque signal is used to infer the power consumption in section \ref{Powsca}.

\subsection{\label{typical}A typical experiment : a complete set of measurements}

\begin{figure}[h]
\centerline{{\includegraphics[width=1\linewidth]{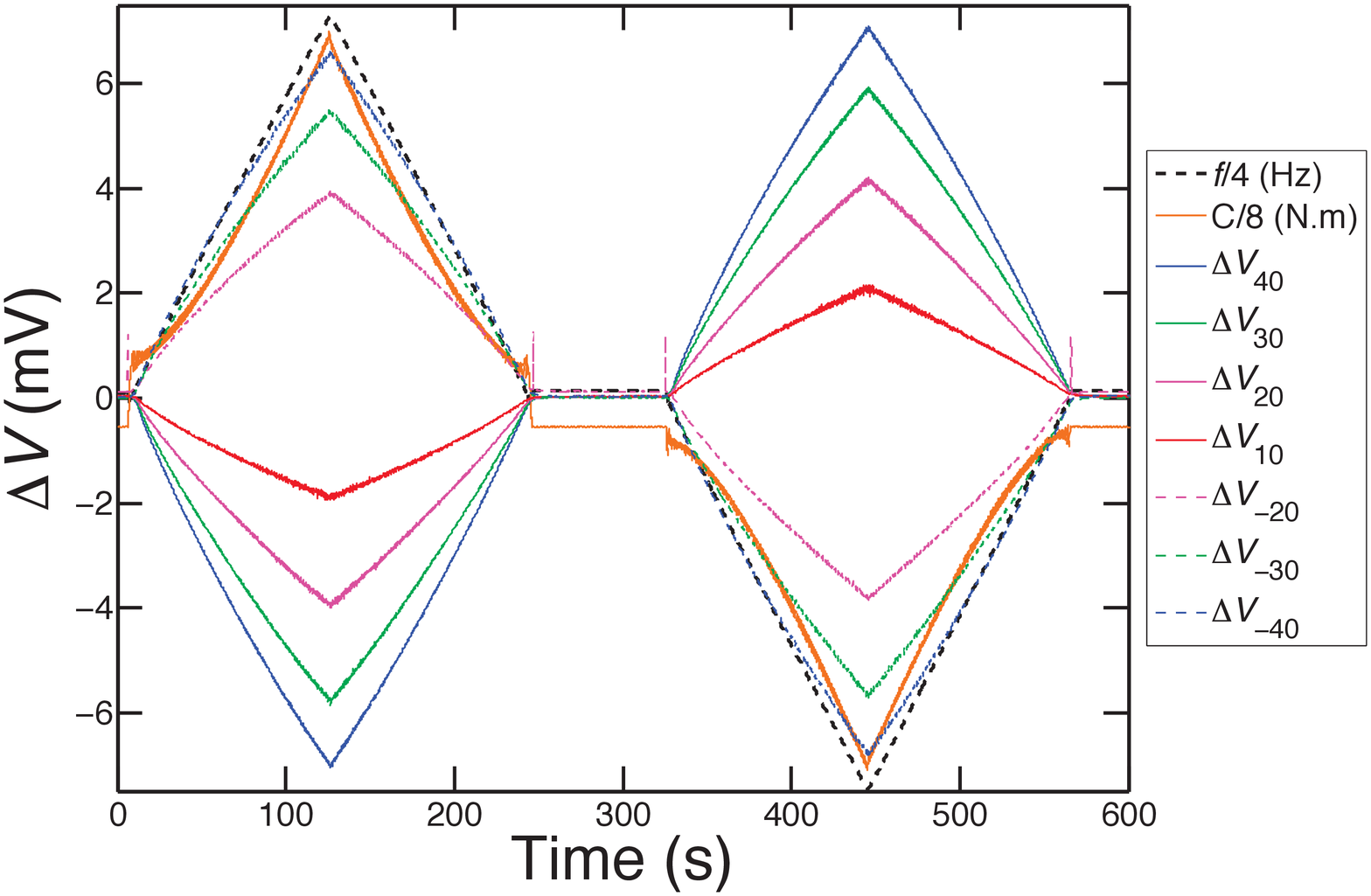}}}
\caption{(Color online) Records of the inner core rotation frequency $ f$, torque C and differences in electric potential $\Delta V_{40}$, $\Delta V_{30}$, $\Delta V_{20}$, $\Delta V_{10}$, $\Delta V_{-20}$,  $\Delta V_{-30}$, $\Delta V_{-40}$ as a function of time. The subscript denotes the latitude (in degrees) of the electric potential difference.}
\label{ElectrodesPaper}
\end{figure}

A complete set of measurements performed during a typical experiment is analyzed below.
The run was chosen to illustrate the various measurements but also to depict how the different observables evolve with $f$.
During that run of 600 seconds,
the inner sphere was first  accelerated from 0 to $ f=30$ Hz in around 120 seconds, then decelerated back to 0 during 120 seconds. The inner sphere was then kept at rest for
about 100 seconds and accelerated in the opposite direction to $ f=-30$ Hz in 120 seconds. It returned to zero rotation in 120 seconds again. That cycle of rotation is shown in FIG.~\ref{ElectrodesPaper}. The torque delivered by the inner sphere motor is also shown and evolves clearly non-linearly during those cycles.

FIG.~\ref{ElectrodesPaper} shows electric potential records (see part \ref{electrodes}) obtained during this experiment and time averaged over 0.1 s windows. The differences of potential vary in sync with the inner sphere rotation frequency as expected if the various $\Delta V$'s measure the differential rotation between the liquid sodium and the outer sphere to which the electrodes are affixed (\ref{electrodes}). However, it is also apparent that the fluid rotation as measured from the $\Delta V$'s does not increase linearly
with the inner sphere frequency. We interpret it as an indication that braking at the outer boundary, which opposes the entrainment by the inner core rotation, varies non linearly with the differential rotation.
As expected, records from electrodes pairs are anti-symmetrical with
respect to the equator, since the forcing is symmetrical while the radial component of the
imposed magnetic field changes sign across the equator.

\begin{figure}[h]
\centerline{{\includegraphics[angle=270,width=0.49\linewidth]{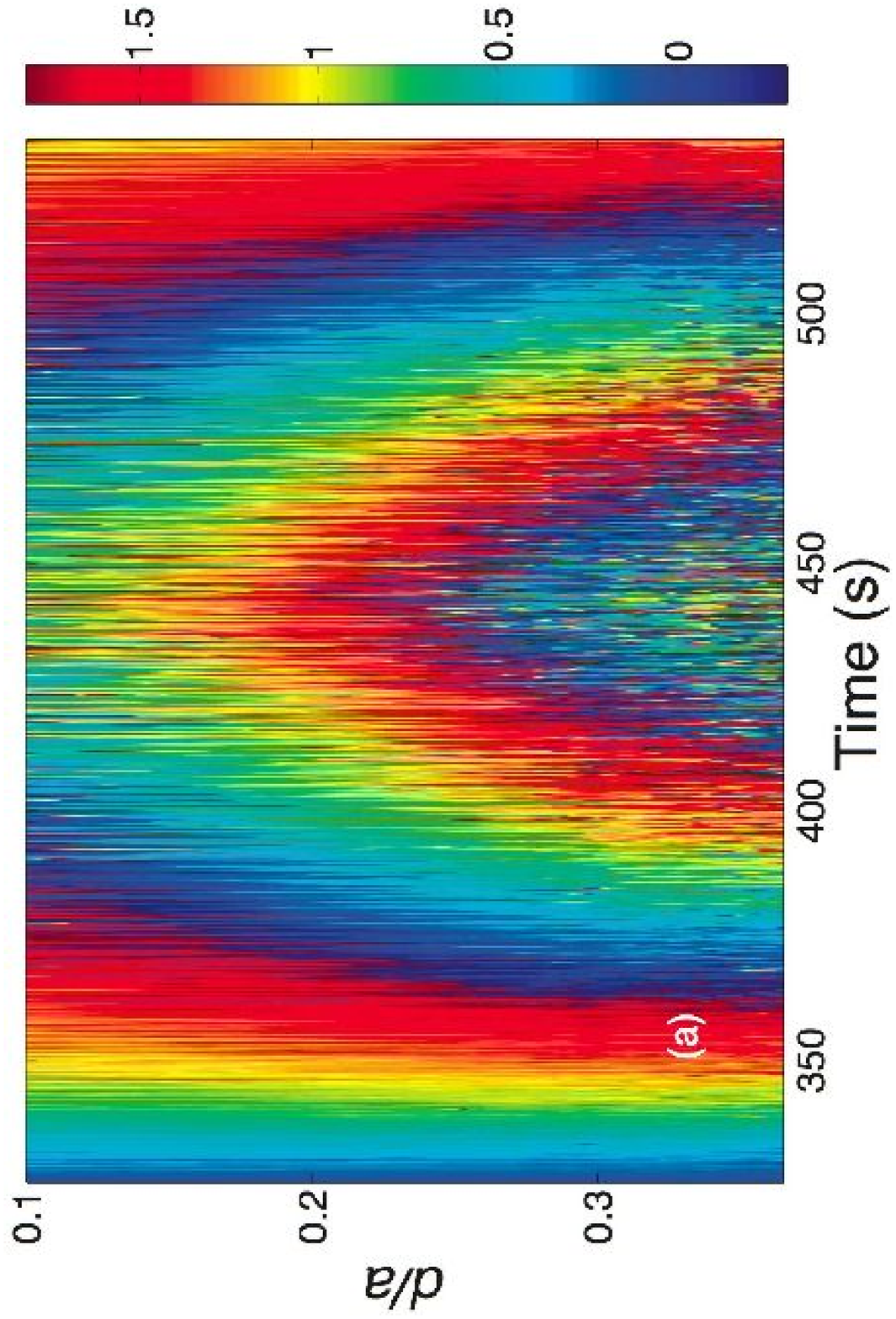}}{{\includegraphics[angle=270,width=0.48\linewidth]{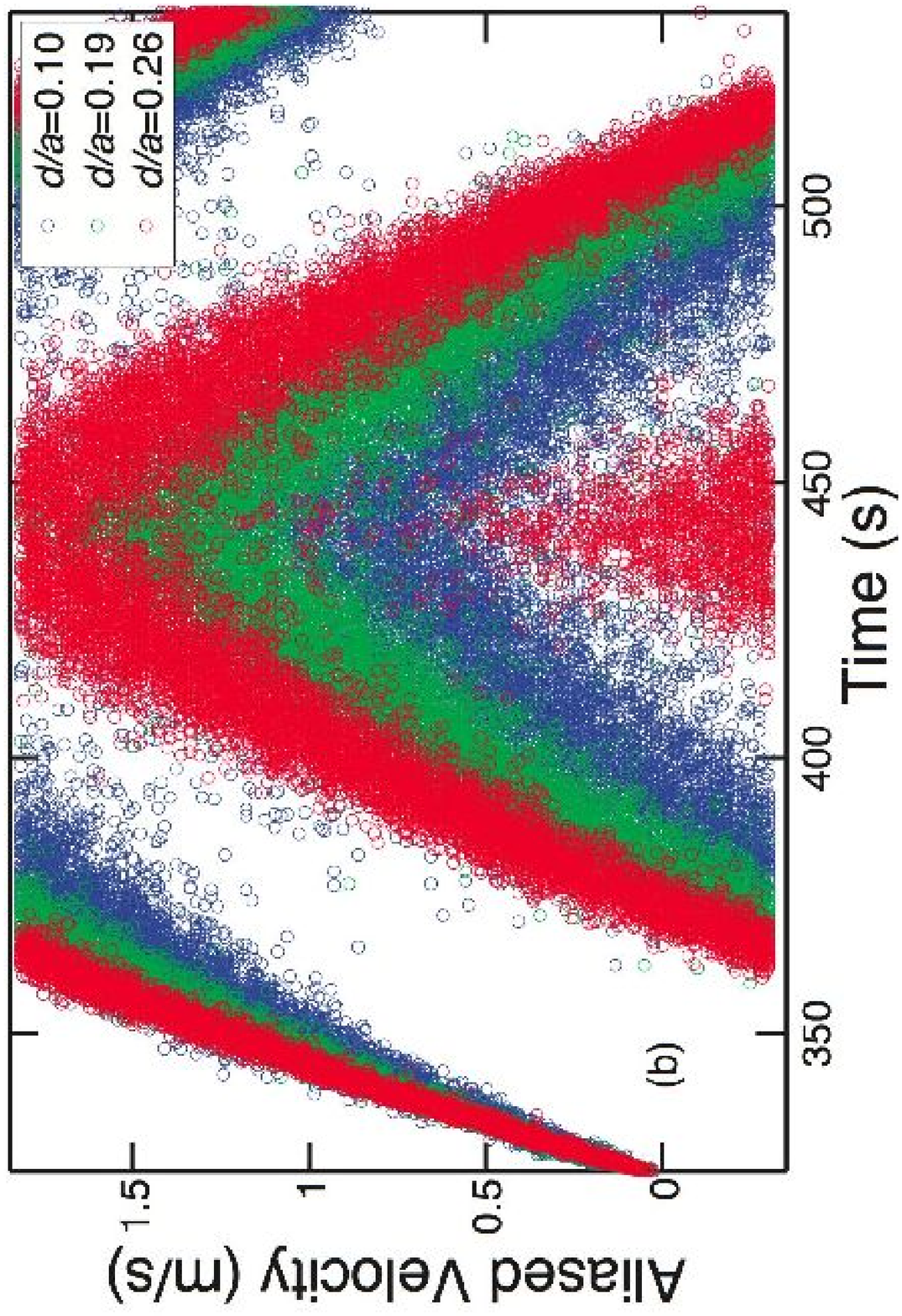}}}}
\centerline{{\includegraphics[angle=270,width=0.90\linewidth]{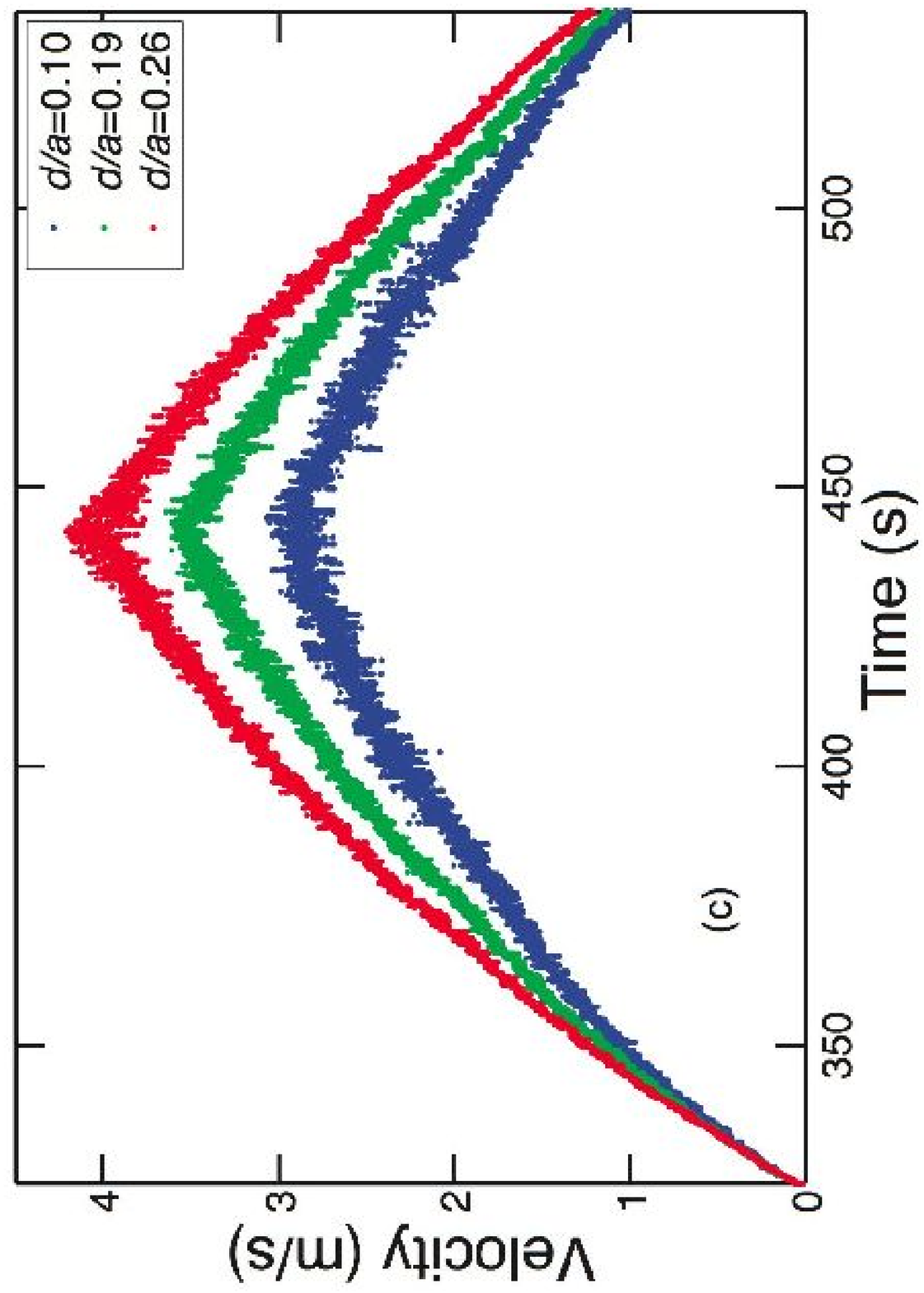}}}
\caption{(Color online) UDV measurements performed along the ray number 6 (see FIG.~\ref{trajectoires}) during the second half of the typical experiment when the inner sphere was rotated from rest to -30 Hz and then back to rest. (a) Spatio-temporal representation of the measured velocity, given by the color scale (in m/s).
(b) Velocity at three distances from the probe as a function of time,  extracted from the spatio-temporal shown in (a). The velocity profiles are clearly aliased since the profiles are discontinuous. (c) After applying a median time-filtering window of 0.2 s and unfolding the profiles, the correct velocities are retrieved as a continuous function of time. } \label{VitessePaper}\end{figure}

FIG.~\ref{VitessePaper} shows the fluid velocity $u(d)$ measured by UDV during the first half of the experiment along the ray 6
as a function of time and distance.
Velocity profiles were recorded along a total distance $d\simeq$ 80 mm.
As demonstrated in FIG.~\ref{VitessePaper}(b), the velocity is aliased since the maximum measurable velocity, for the ultrasonic frequency used during the experiment, is exceeded. Since the azimuthal velocity profiles are quite simple in shape, it has been straightforward to unfold those profiles and retrieve the correct amplitudes as shown in FIG.~\ref{VitessePaper}(c).
The evolution with $f$ is similar to that of the electrodes, but indicates a stronger leveling-off as $f$ increases.

\begin{figure}[h]
\centerline{{\includegraphics[width=0.50\linewidth]{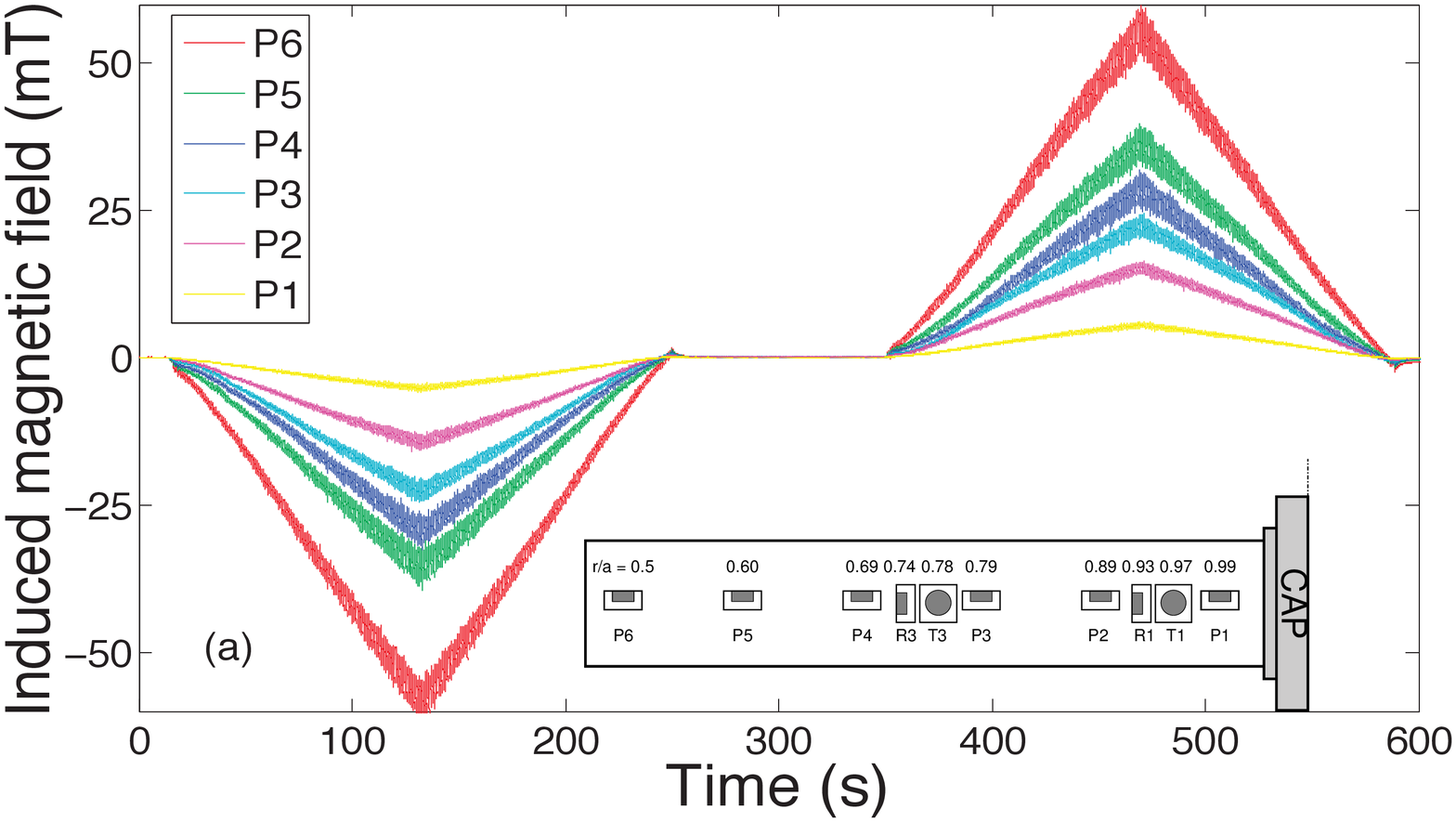}}{\includegraphics[width=0.485\linewidth]{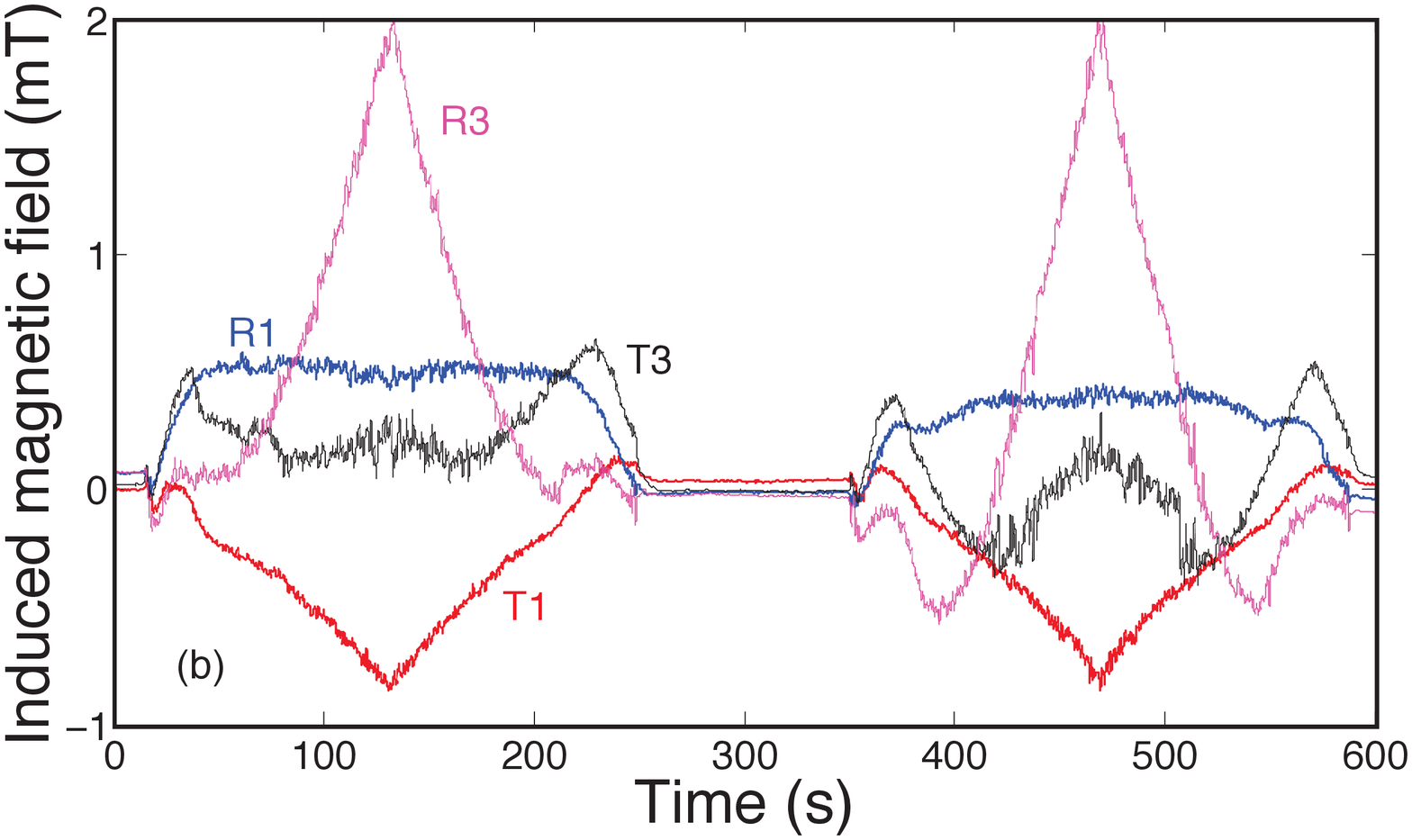}}}
\caption{(Color online) (a) Azimuthal $b_{\varphi}$, (b) radial $b_r$ and orthoradial $b_{\theta}$ induced magnetic field  at a latitude
of $40\deg$ in the sleeve at different radial positions recorded during the two triangles sequence of FIG.~\ref{ElectrodesPaper}. A top view of the sleeve at the bottom of (a) gives the radial position and the orientation of the various Hall magnetometers. The intensity of the induced azimuthal field reaches $60$ mT near the inner sphere
and has the sign of $-f$. The
fluctuations reach about 10$\%$ of the mean. The meridional components of the induced magnetic field are
much weaker and dominated by fluctuations, which have been filtered out here ($0.2$ Hz low-pass filter).}
 \label{inducedB}\end{figure}

FIG.~\ref{inducedB} shows the magnetic field induced inside the fluid during the typical experiment. The
measurements are taken in the sleeve placed at $40\deg$ latitude. The induced azimuthal field in FIG.~\ref{inducedB} (a) is
measured at 6 different radii (given in section \ref{doigt_de_gant}). Its intensity reaches $60$ mT near the inner 
sphere and gets larger than the imposed dipole in some locations. Note the simple evolution with $f$, which contrasts
with that of the electric potentials and velocities in that it increases with an exponent close to 1. The induced 
meridional field (FIG.~\ref{inducedB}) is some 20 times weaker. It is dominated by fluctuations, and does
not change sign when $f$ does. Note that the evolution with $f$ is not monotonic.
Similar behaviors are observed at latitudes $10\deg$ and $-20\deg$.

\clearpage

\subsection{Power scaling}
\label{Powsca}

\begin{figure}[h]
\centerline{\includegraphics[clip=true,width=1\linewidth]{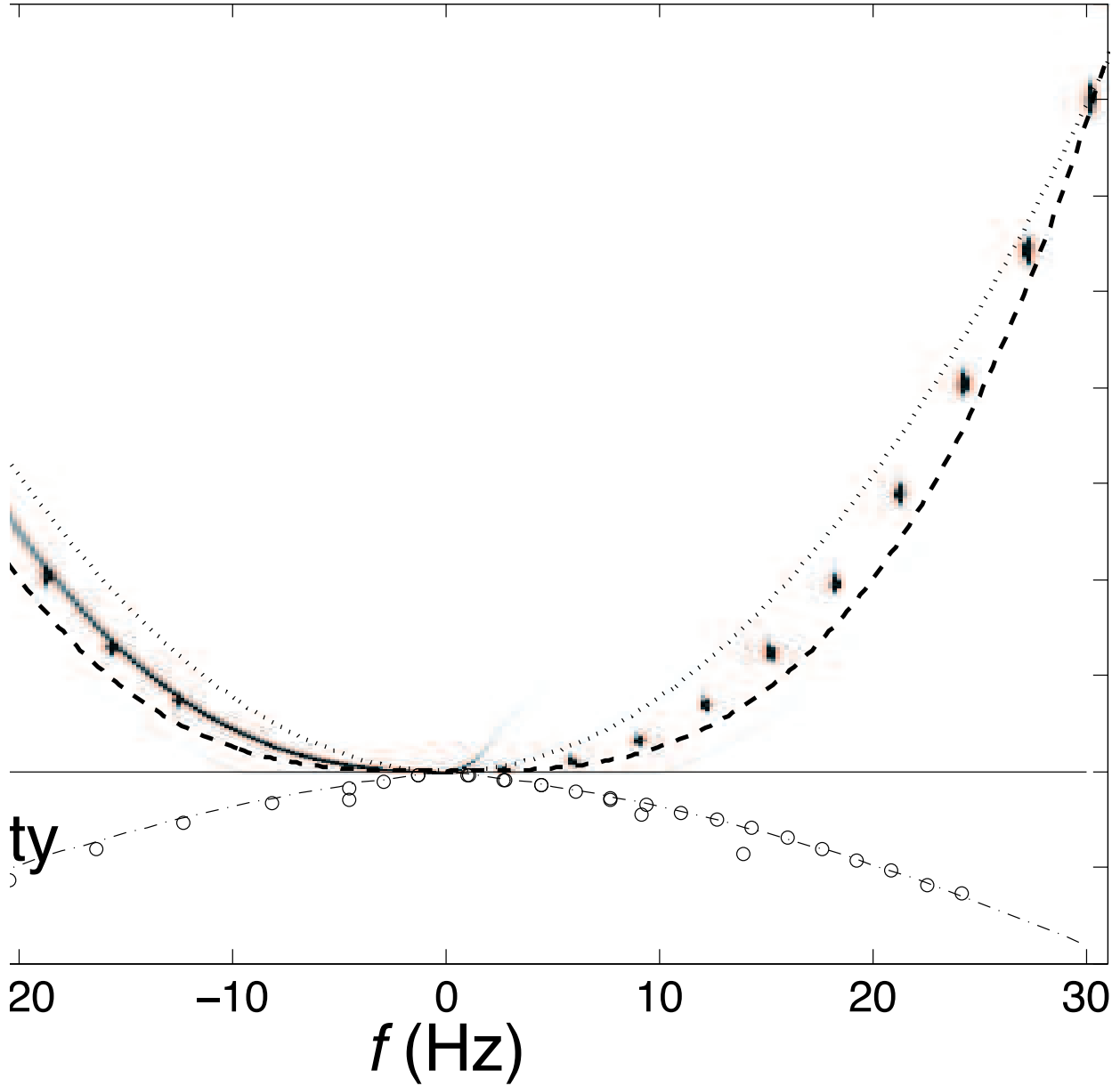}}
   \caption{Power dissipated by the flow in $DTS$. The data dots are from measurements of the motor torque for
   plateaus at given $f$. The dissipation in the mechanical set-up has been removed. It is obtained by rotating the
   inner sphere before
   filling the shell with sodium.
   It is drawn here upside-down in the lower panel (empty symbols) and can be fit by
   ${\cal{P}}_{\mbox{\tiny{empty}}} (\mbox{W}) = 4 \times | 2\pi f | + 0.03 \times (2 \pi f)^2$ (dash-dot curve).
   Dissipation in the flow scales as $f^{2.5}$, and is here compared with $f^2$ (dotted line) and $f^3$ (dashed line).
}
   \label{power}
   \end{figure}

The power dissipated by the flow is shown in FIG.~\ref{power}
as a function of the rotation
frequency $f$. It is computed from the product $\Gamma$ $\times$ $2 \pi  f$, where $\Gamma$ is
the torque retrieved from the motor drive. We subtracted the power measured with an empty shell (dash-dot curve) to eliminate
power dissipation in the mechanical set-up. The dissipation in the fluid reaches almost $8$ kW for the highest
rotation frequency of the inner sphere ($f=\pm30$ Hz). The small spread of the data dots indicates that power
fluctuations are small. The continuous line is the record of power versus $f$ when the inner sphere is ramped from
$0$ to $-30$ Hz as in FIG.~\ref{ElectrodesPaper}. The corresponding increase in kinetic energy only slightly
augments power dissipation.

Power dissipation is found to scale as $f^{2.5}$, which does not differ from the scaling obtained in the laminar numerical study of section \ref{comparison}. There, it is explained as the result of the balance between the magnetic torque on the inner sphere and the viscous torque on the outer sphere, assuming that the fluid angular velocity below the outer viscous boundary layer is of the order of the inner sphere angular velocity. Although the outer boundary layer displays strong fluctuations, the situation is completely different from  Taylor-Couette water experiments \cite{lathrop1992turbulent}.

\clearpage
\section{Governing equations}
\label{equations}

A spherical shell of inner radius $b$ and outer radius $a$ is immersed in an axisymmetric dipolar magnetic field $\mathbf{B_d}$: 
$$\mathbf{B_d}(r,\theta, \varphi) = B_0\left (\frac{b}{r}\right)^3\left [ 2  \cos\theta \mathbf{e_r} +
\sin\theta \mathbf{e_{\theta} }\right ], $$ where $(r,\theta,\varphi)$ are spherical coordinates.
The outer boundary is kept at rest and the inner sphere rotates with the constant angular velocity $\Omega=2\pi f$ along the same axis as the dipole field that it carries. We assume that the electrically conducting fluid filling the cavity is homogeneous, incompressible and isothermal. We further assume that the flow inside the cavity is steady. 

The inner body consists of a magnetized innermost core enclosed in an electrically  conducting spherical solid envelope of finite thickness $d_b$.  We choose $b$ as unit length, $b\Omega$ as unit velocity, $\rho b^2 \Omega^2$ as unit pressure, and $b^2\Omega B_0/\eta = \Rm B_0$ as unit of induced magnetic field $\mathbf{b}$ ($\mathbf{B}= \mathbf{B_d}+\Rm\mathbf{b}$). Then, the equations governing the flow $\mathbf{u}$ and the induced magnetic field are:

\begin{eqnarray}
\mathbf{\nabla} \cdot \mathbf{u} &=&0 \label{divu}\\
\mathbf{\nabla} \cdot \mathbf{b} &=&0 \label{divb}\\
(\mathbf{u}\cdot \mathbf{\nabla})\mathbf{u}&=& -\nabla p + \Lambda  \left((\mathbf{B_d}\cdot \mathbf{\nabla})\mathbf{b}+(\mathbf{b}\cdot \mathbf{\nabla})\mathbf{B_d}
\right)+ \Re^{-1}\mathbf{\nabla}^2\mathbf{u}
\label{motion}\\
\mathbf{\nabla}^2 \mathbf{b} &=& -  \mathbf{\nabla}\times (\mathbf{u}\times\mathbf{B}),\label{induction}
\end{eqnarray}
where $p$ is a modified pressure. The notation $\Lambda$ refers to the Elsasser number, classically used for rotating flows in the presence of a magnetic field. That number $\Lambda$ compares the magnetic and inertial forces in the vicinity of the magnetized inner sphere. In the shell interior, the two forces are better compared by a "local" Elsasser number: $\Lambda_l = (b/r)^6\Lambda$ (with $(b/a)^6\simeq 1.83 \,10^{-3}$). Finally, it is of interest to introduce the Hartmann number $\Ha$ that compares the magnetic and viscous forces. We have $\Ha=(\Lambda \Re)^{1/2}$. In the shell interior, the number $(b/r)^3 \Ha$ is more appropriate to compare the two forces. Typical values of these dimensionless numbers can be found in TABLE~\ref{sansdim}.

\begin{table}
\caption{\label{sansdim}Typical values of the dimensionless numbers in the $DTS$ experiment, computed for $f=\Omega/2 \pi = 25$ Hz. }
\begin{ruledtabular}
\begin{tabular}{ccc}
 
 $\Re$ & $b^2\Omega/ \nu$ & {$1.3\;10^6$}\\
 $\Rm$ &  $b^2\Omega/ \eta$
 &{10} \\
 $\Lambda$ &  $\sigma B_0^2/\rho \Omega$ &  1.9\\
 $\Ha$ & $(\Re\Lambda)^{1/2}$ & $1.6\, 10^{3}$
\end{tabular}
\end{ruledtabular}

\end{table}

The set of equations (\ref{divu}-\ref{induction}), where the non linear terms are neglected, was the subject of the analytical study of Dormy et al. \cite{dormy02} that described how the differential rotation between the fluid interior and the outer sphere
drives an influx of electrical currents from the mainstream into the outer viscous Hartmann boundary layer.
Electrical currents flow along the viscous boundary layer and return to the conducting inner body along a free shear
layer located on the magnetic field line tangent to the outer boundary at the equator. 
As these electrical currents
cannot flow exactly parallel to the magnetic field line, they produce a Lorentz
force, which sustains "super-rotation" of the fluid. 
Recent studies have extended the analysis to the case of a finitely conducting outer sphere \cite{mizerski2007effect,soward2010shear}.
On increasing the conductance of the container, Dormy et al. (2010) found that more and more electrical currents leak into the solid boundary and the super-rotation rate
gets as large as $O(\Ha^{1/2})$. 
Though the analytical results have set the stage for the interpretation of the experimental results, the neglected non linear effects are crucial in the $DTS$ experiment, even for the smallest rate of rotation of the solid inner body.

Upon reversal of $\Omega$, $u_\varphi$ and $b_\varphi$ change into $- u_\varphi$ and $-b_\varphi$ whilst the other components of $\mathbf{u}$ and  $\mathbf{b}$ are kept unchanged.

\clearpage
\section{Differential rotation}

\subsection{Transition between the Ferraro and geostrophic regimes}
\label{Ferraro}

In this section, we use the UDV records to delve into the geometry of isorotation surfaces.

The L number associated to each dipolar magnetic field line enters the equation of the surfaces spanned by dipolar lines of force:
\begin{equation}
r = \mbox{L} \sin^2  \theta \; .
\end{equation}

\begin{figure}[h]
\centerline{{\includegraphics[width=0.52\linewidth]{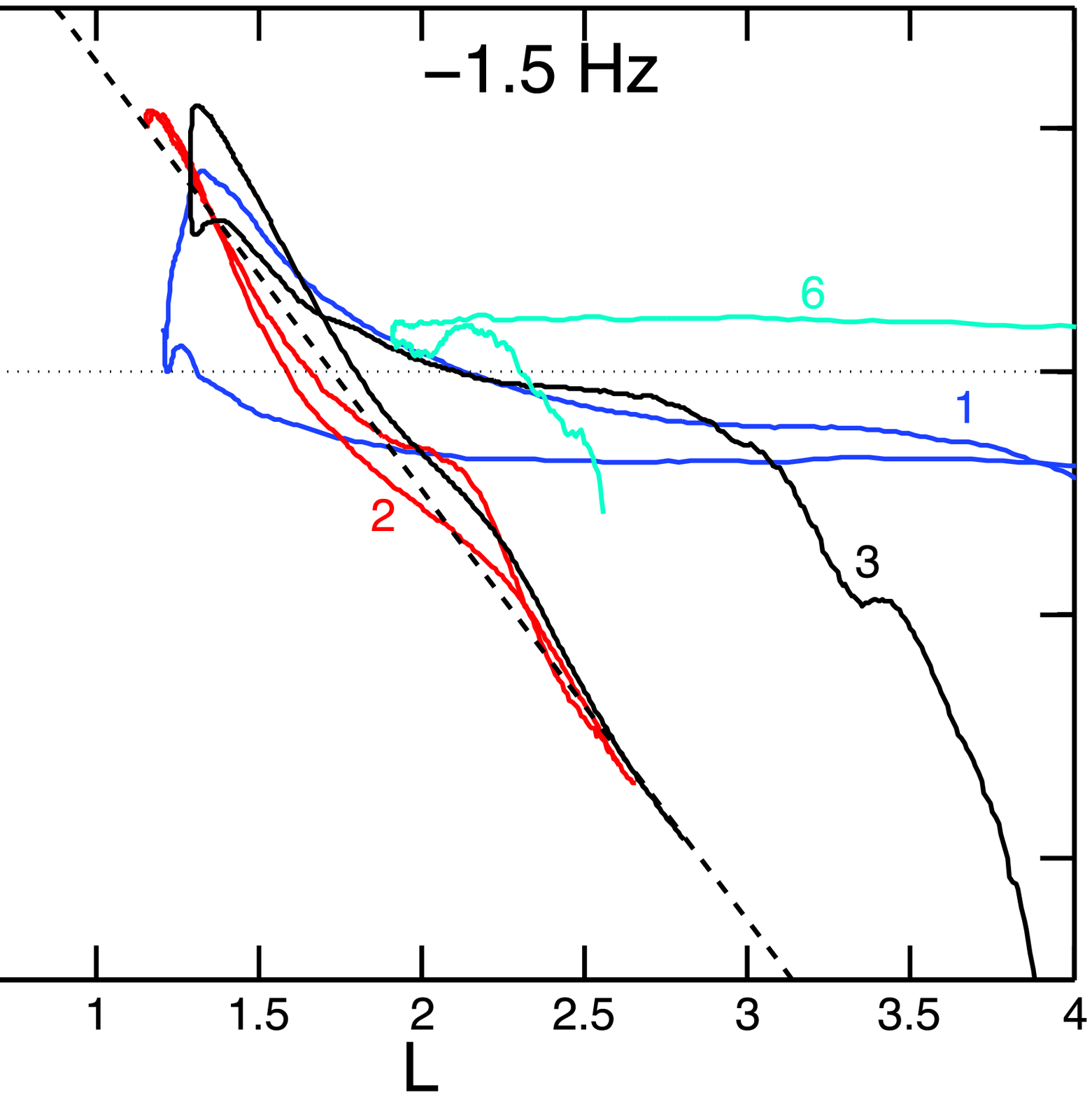}}{\includegraphics[width=0.52\linewidth]{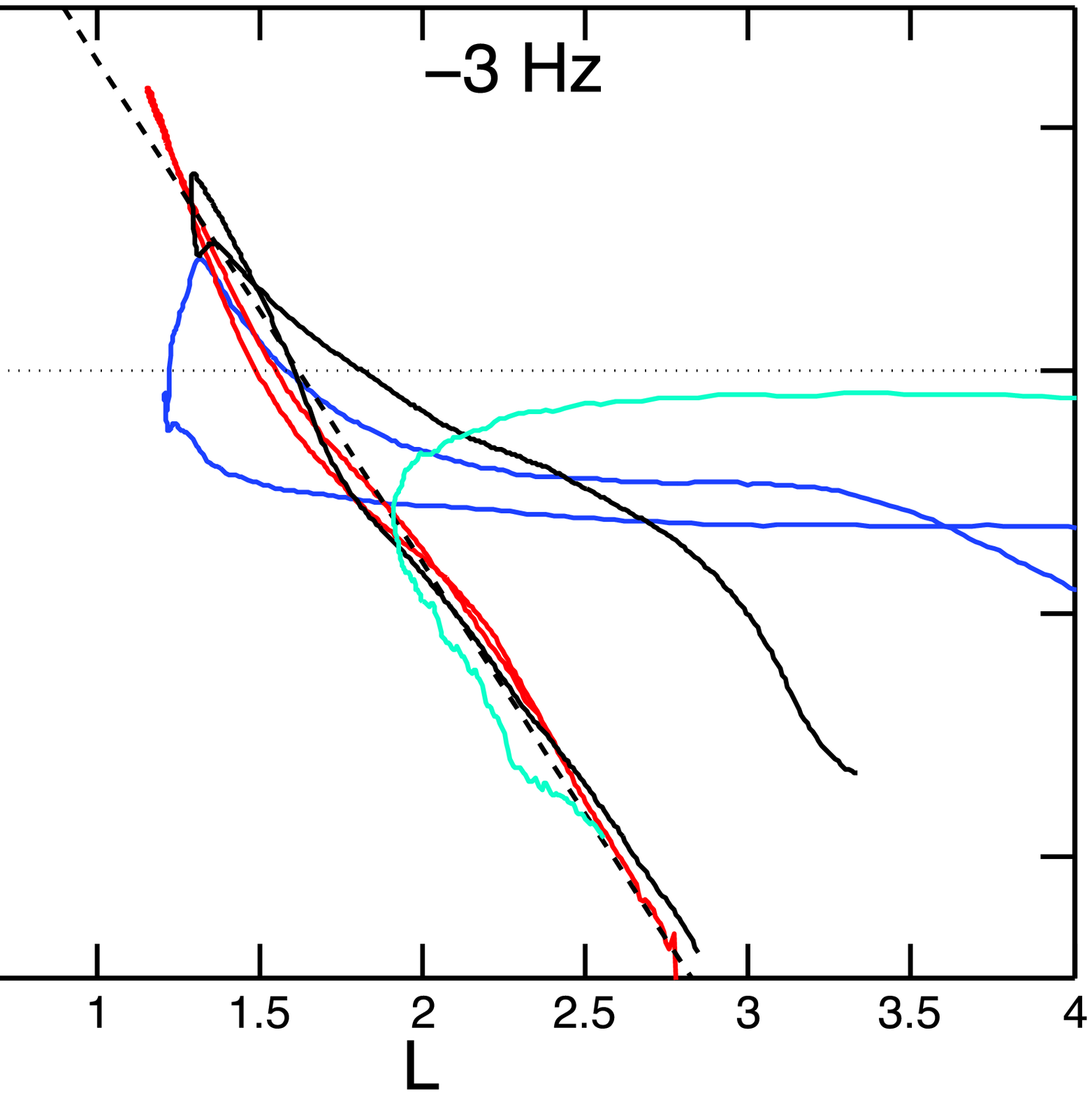}}}
\centerline{{\includegraphics[width=0.52\linewidth]{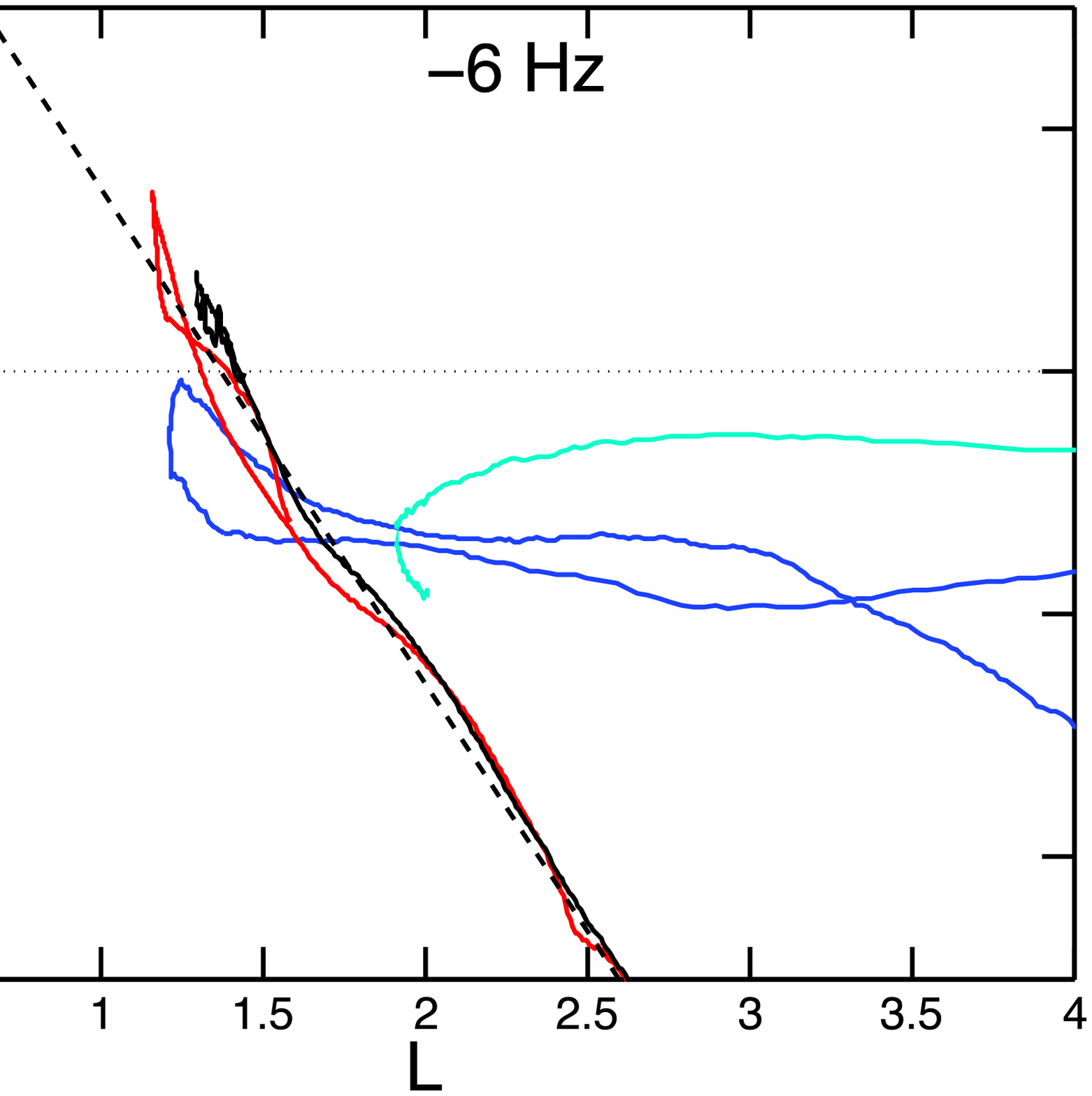}}{\includegraphics[width=0.52\linewidth]{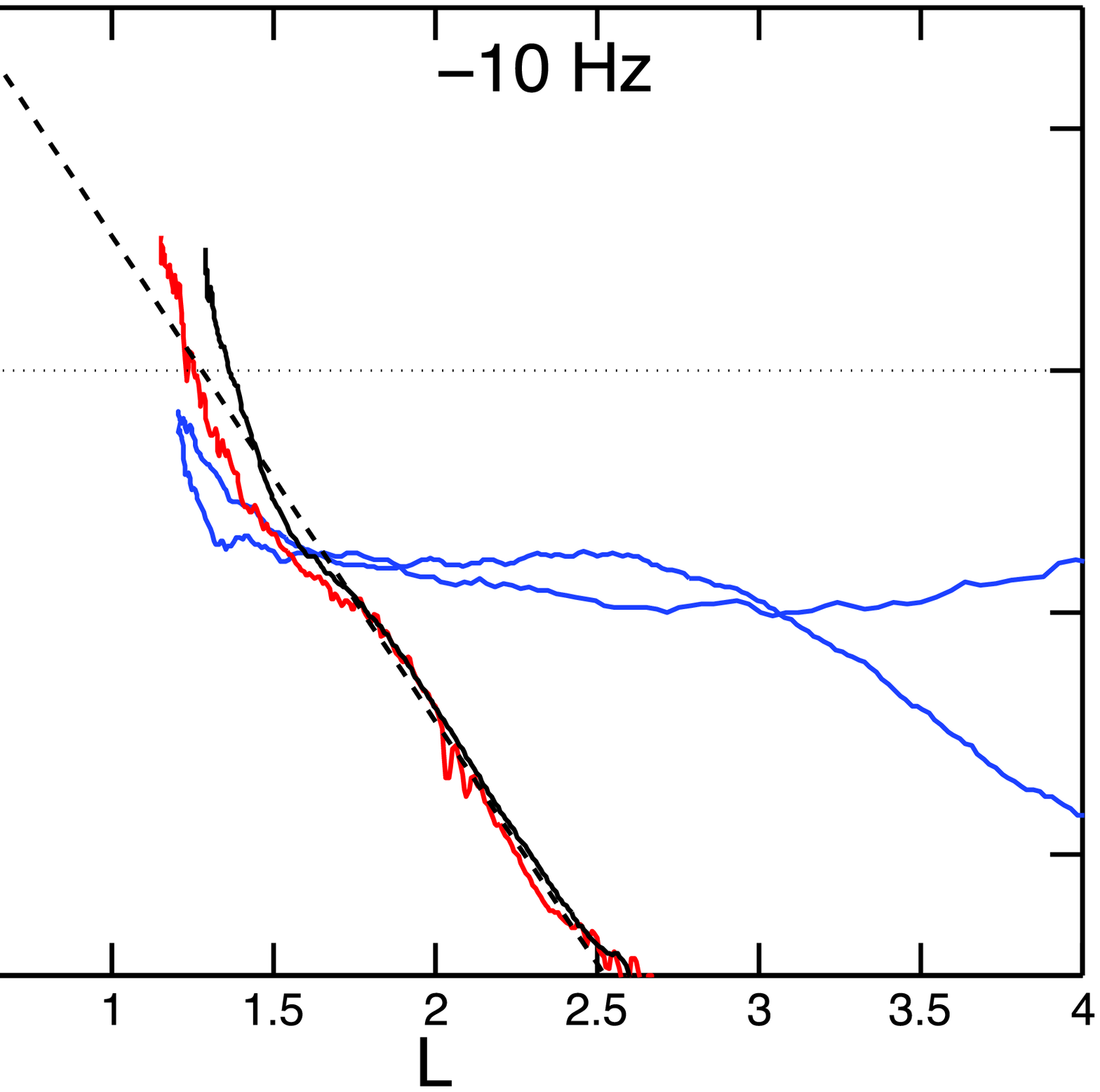}}}
   \caption{(Color online) 
   Rotation frequency of the fluid sodium over the inner sphere rotation frequency as a function of the magnetic field lines L for four ultrasonic velocity profiles (trajectories 1, 2, 3 and 6, with the same color code as in FIG~\ref{trajectoires}) and four inner sphere rotation frequencies ($f=$ -1.5, -3, -6 and -10 Hz).
The dashed line is a straight line to help the eye.}
\label{ferraro}
\end{figure}

Accordingly, $L$ gives the radius of the intersection of the magnetic field line with the equatorial plane.
The notation L refers to the L-value (or L-shell parameter) widely used to describe motions of low energy particles in the Earth's magnetosphere.
FIG.~\ref{ferraro} shows that, for L$\leq 2.7$, the angular velocity measured along rays 2 and 3, which are the most appropriate to map the azimuthal velocity field, is, to a large extent, a function of L only. Thus, the angular velocity does not vary along  magnetic field lines near the inner sphere, where the magnetic field is the strongest. We interpret this result as a consequence of Ferraro's theorem of isorotation. The latter is written:
\begin{equation}
\mathbf{B_d}\cdot\nabla\left(\frac{u_\varphi}{s}\right)=0.
\label{eq:Ferraro}
\end{equation}
\noindent
It is obtained from the $\varphi$ component of the induction equation for steady fields, ignoring magnetic diffusion. 
Although often invoked in the framework of ideal MHD (where magnetic diffusion is negligible), Ferraro's law does not
require a large $\Rm$ \cite{allen1976law}.
It implies that there is no induced magnetic field and that, as a consequence, the magnetic force is exactly zero. 
More precisely, deviations from this law lead to the induction of a magnetic field, which produces a magnetic force that tends to oppose this induction process.
Writing $\mathbf{u}=\mathbf{u_0}+\mathbf{u_1}$, where $\mathbf{u_0}$ obeys the equation (\ref{eq:Ferraro}), we obtain $b\approx u_1$ from (\ref{induction}). Then, the momentum equation (\ref{motion}) yields $u_1\approx (\Re\Lambda)^{-1} u_0= \Ha^{-1} u_0$ (as numerically verified in \cite{macgregor1999angular}) when the inertial term, on the left hand side, can be neglected.
Ferraro's law of isorotation, though, is not the only way to cancel the magnetic force. In the presence of electric currents parallel to the magnetic field, the magnetic force remains zero and the equation (\ref{eq:Ferraro}) can be violated \cite{allen1976law,soward2010shear}. For the geometry of the $DTS$ experiment, it cannot happen along the innermost dipolar field lines that join the two hemispheres, without touching the outer sphere. Indeed, symmetry with respect to the equatorial plane E implies that the currents do not cross E.

Thus, the observation
of a velocity field obeying Ferraro's law is a symptom that magnetic forces predominate in that region.
Note that the fact that the two legs of the profile along ray 2 show similar velocities even for large L only probes the symmetry of the flow with respect to the equatorial plane.

\begin{figure}[h]
 \centerline{{\includegraphics[width=0.50\linewidth]{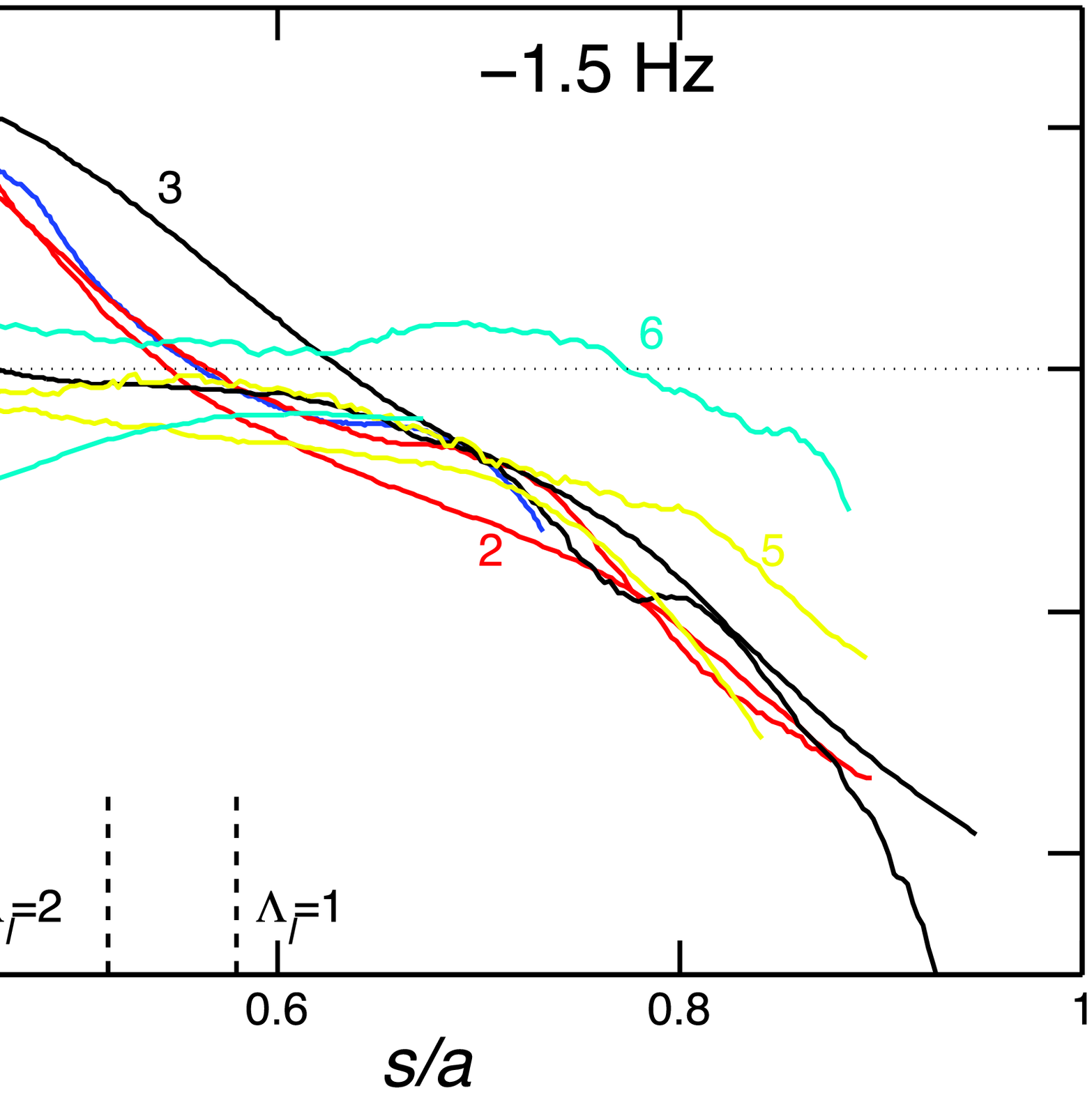}}{\includegraphics[width=0.50\linewidth]{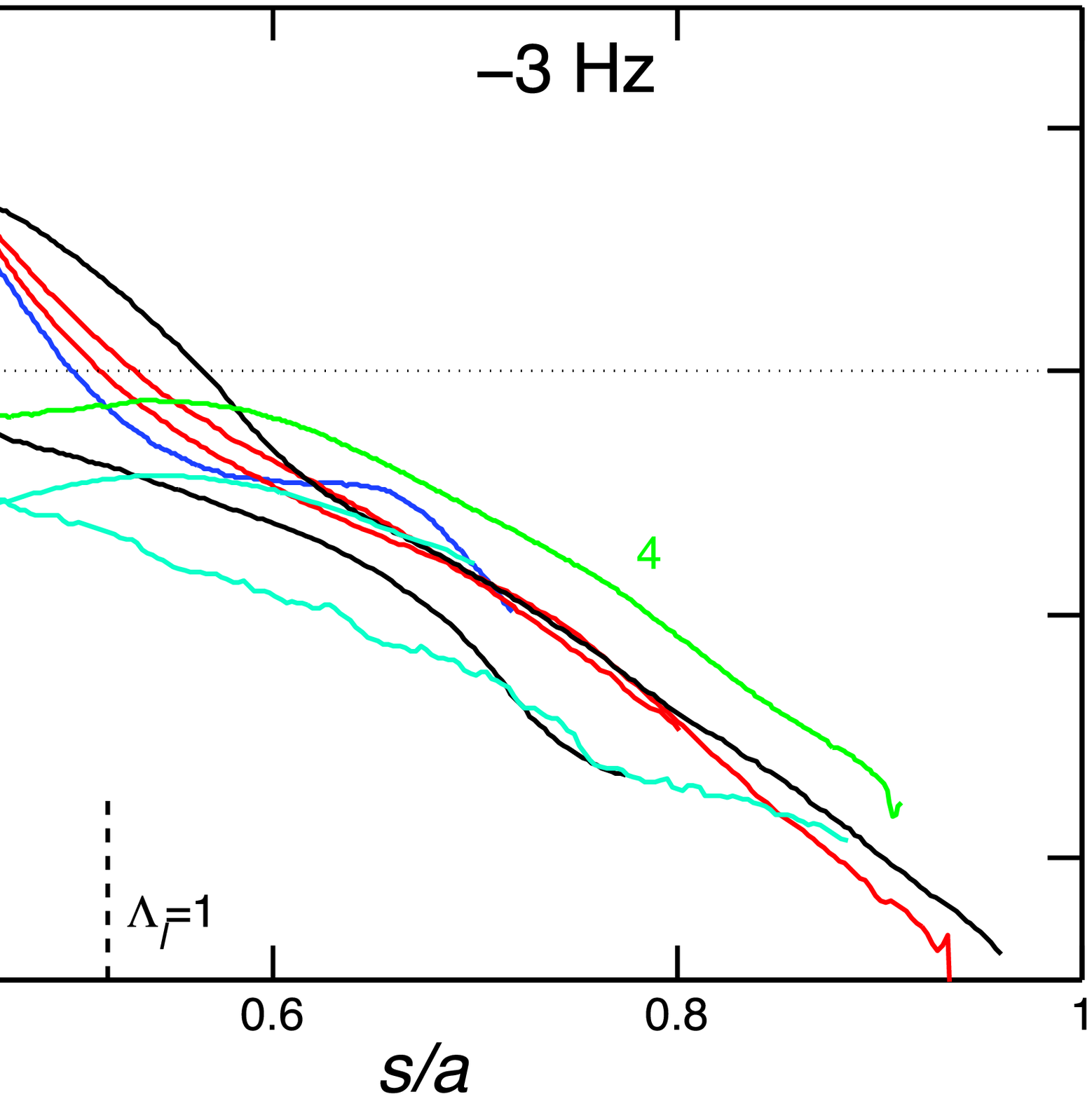}}}
 \centerline{{\includegraphics[width=0.50\linewidth]{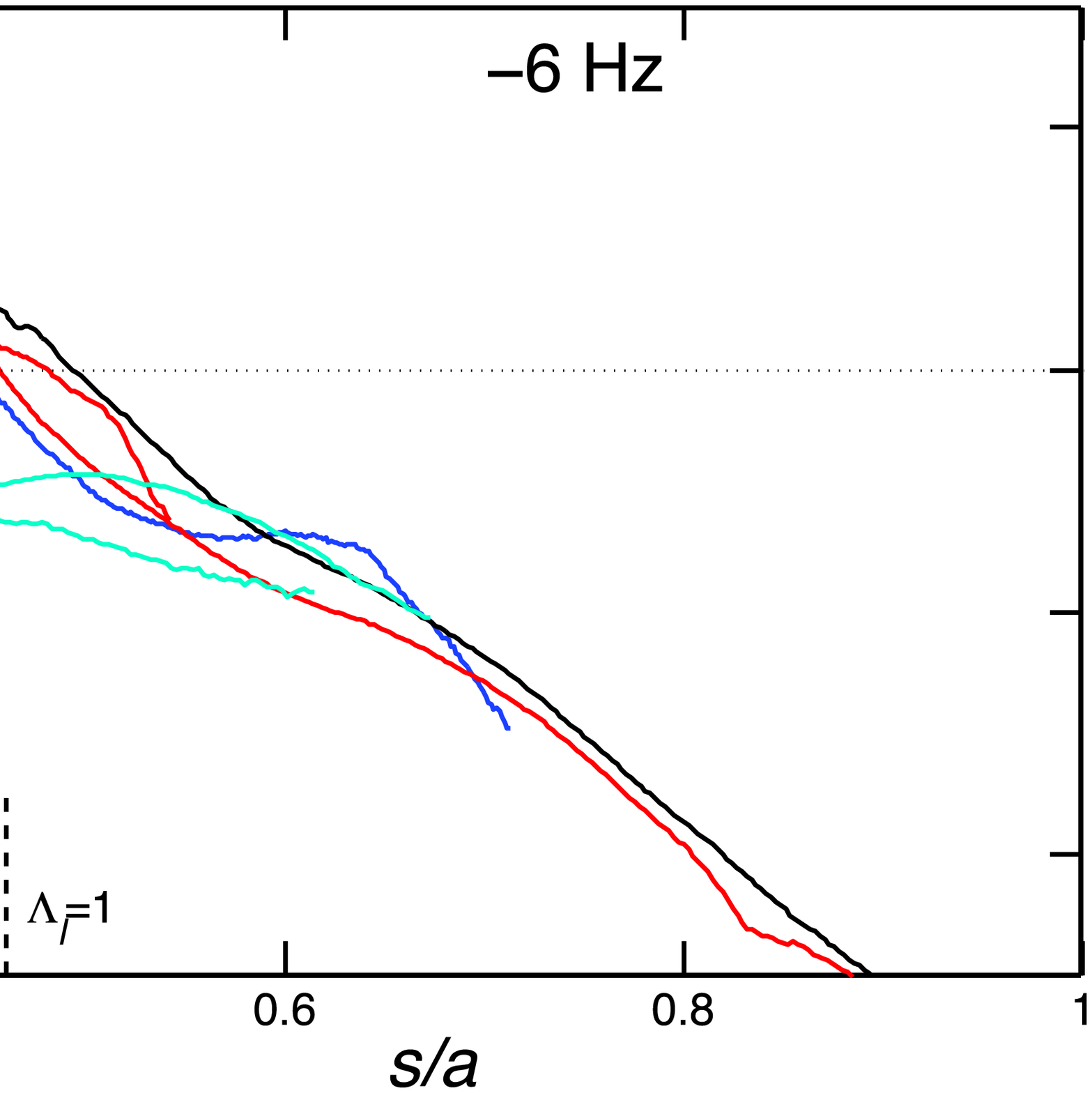}}{\includegraphics[width=0.50\linewidth]{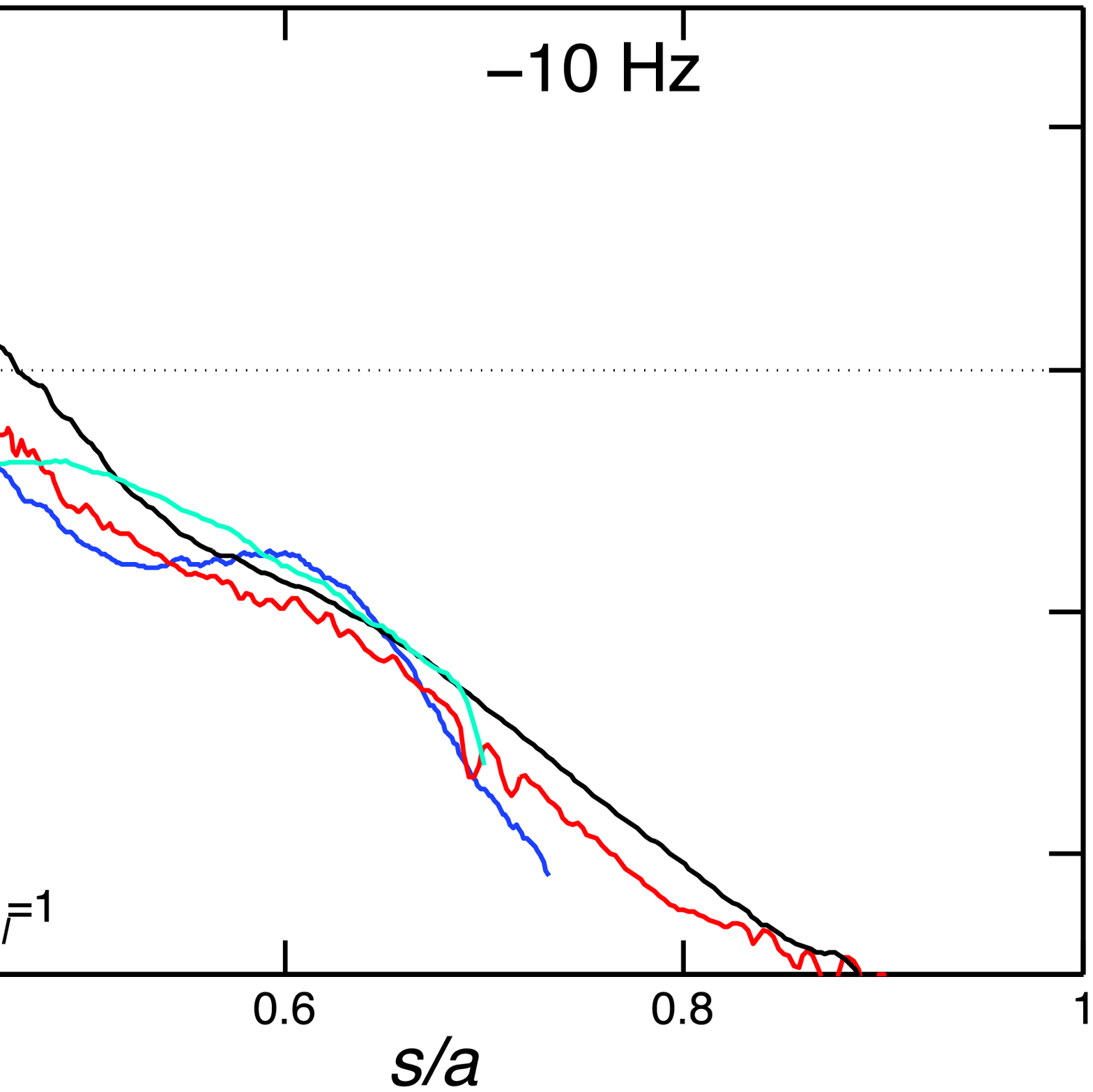}}}
   \caption{(Color online) Rotation frequency of the fluid sodium normalized by the inner sphere rotation frequency as a function of $s$,  for various ultrasonic  velocity profiles and four inner sphere rotation frequencies ($f=$ -1.5, -3, -6 and -10 Hz). The colors of the profiles (numbers) follow the conventions laid out in  FIG.~\ref{trajectoires}.}
   \label{geostrophy}
\end{figure}

Now, FIG.~\ref{geostrophy} shows that for $s\geq0.6$ the azimuthal velocity is largely a function of $s$ only. There, the Proudman-Taylor theorem holds and azimuthal flows are geostrophic as the inertial forces predominate. In contrast with the case of a rotating outer sphere (see Figure 7 in \cite{nataf07}), there is no region of uniform rotation: zonal velocities are $z$-independent but vary with the distance to the $z$ axis.

\begin{figure}[p]
 \centerline{\includegraphics[width=1\linewidth]{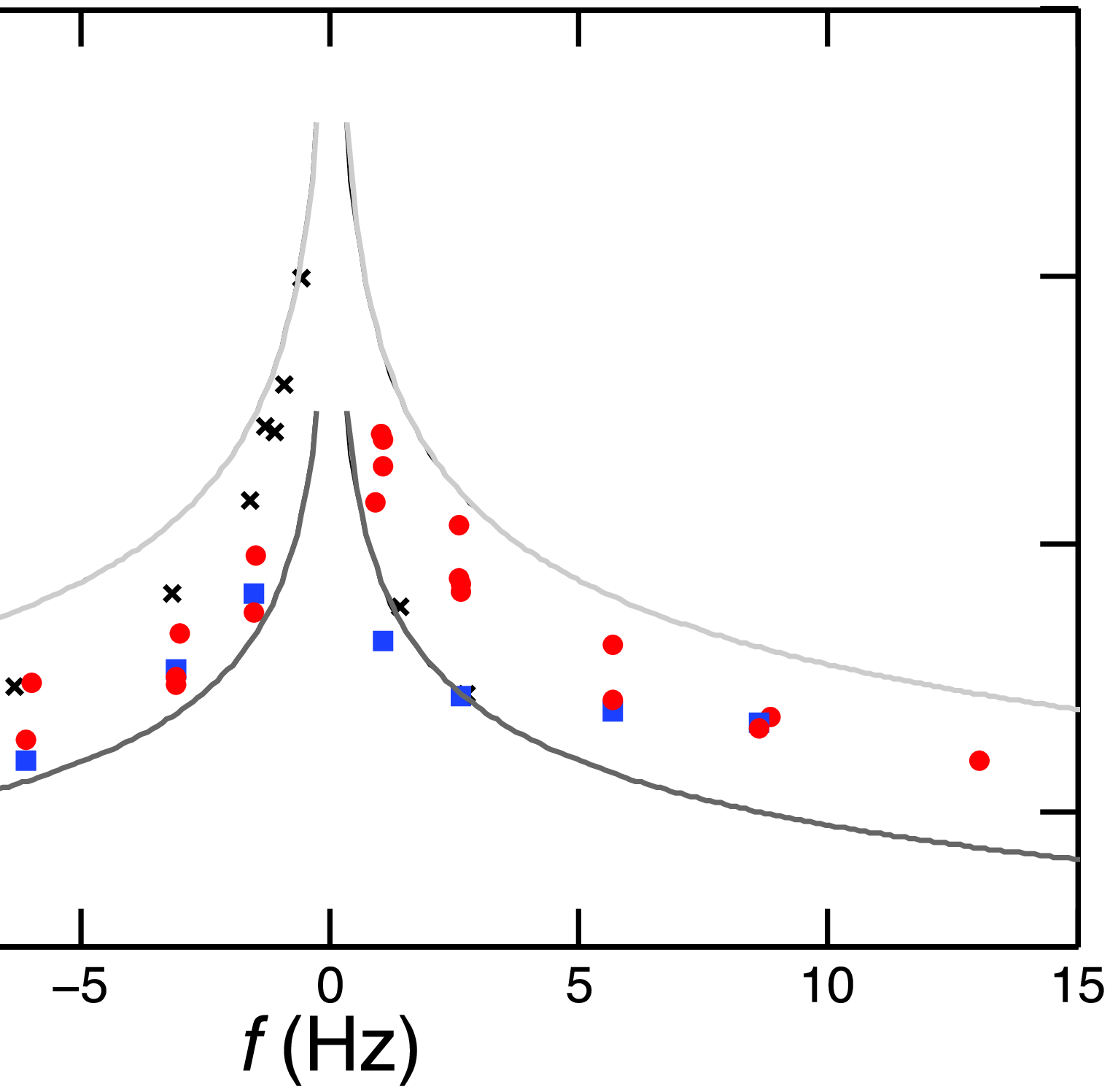}}
   \caption{(Color online) Normalized cylindrical radius $s/a$ along the UDV trajectories number 1 (blue square), 2 (red circle) and 3 (black cross)  where $ f_{fluid} =  f$  ({\it i.e.} $f^*=1$) as a function of the inner sphere rotation frequency. Pale line : $\Lambda_l = 0.5$, Dark line : $\Lambda_l = 2.5$. }
   \label{sVsEls}
\end{figure}

The transition between the Ferraro and geostrophic regimes (FIG.~\ref{sVsEls}) occurs at smaller distances from the axis as the rotation frequency of the inner core increases (unfortunately, we cannot get reliable UDV data for larger $f$). It takes place where the local Elsasser number $\Lambda_l$, which compares the magnetic and inertial forces, is of order 1.
It is noteworthy that the Elsasser number $\Lambda$ defines the location (cylindrical radius) where $\Lambda_l=1$. The surface $\Lambda_l=1$ separates two regions of the fluid cavity. Inside this surface, the magnetic forces predominate whether outside it the rotation forces are the most important ones. Finally, the value of $\Lambda$ largely defines the geometry of isorotation surfaces.

In the geostrophic region, magnetic stress integrated on the geostrophic cylinders remains strong enough to overcome the viscous friction at the outer boundary and to impart a rapid rotation to the fluid but becomes weaker than the Reynolds stress (which can be represented as a Coriolis force). As a result, the fluid angular velocity is still of the order of the angular velocity of the inner sphere and the velocities are predominantly geostrophic.

\subsection{Inversion of velocity profiles}

Flow velocity is constrained by its projection on the several ultrasonic rays that we shoot. We invert the Doppler velocity profiles for the large scale mean flow, assuming that the steady part of the flow is symmetric about the axis of rotation and with respect to the equatorial plane.
A poloidal/toroidal decomposition, 
\begin{equation}
\mathbf{u} = u_\varphi \mathbf{e_\varphi} +\mathbf{\nabla} \times
(u_p \mathbf{e_\varphi})  \; ,
\label{pol-tor}
\end{equation}
is employed.
We first consider the azimuthal velocity $u_\varphi$, which is expanded in associated Legendre functions with odd degree and order 1, {\it i.e.}
\begin{equation}
u_\varphi(r,\theta)=\sum_{l=0}^{l_{max}}u_\varphi^l(r) P_{2l+1}^1(\cos\theta) \; .
\label{odd-leg}
\end{equation}
The functions $u_\varphi^l(r)$ are decomposed into a sum from $k=0$ to $k_{max}$ of Chebyshev polynomials of the second kind on the interval $\left[0,1\right]$ mapped onto the interval $\left[b/a, 1\right]$, {\it i.e.} the fluid domain. The azimuthal velocity is not constrained to vanish at the inner and outer boundaries, in order to account for the presence of thin unresolved boundary layers.

Azimuthal velocities are more than 10 times larger than the poloidal ({\it i.e.} meridional) velocities. Nevertheless, the latter
projects onto the ultrasound rays. We take the difference of the profiles acquired for $f$ and $-f$ in order to eliminate this small contribution (the meridional circulation does not change sign while the azimuthal velocity does).

\begin{figure}[h]
\centerline{{\includegraphics[width=0.40\linewidth]{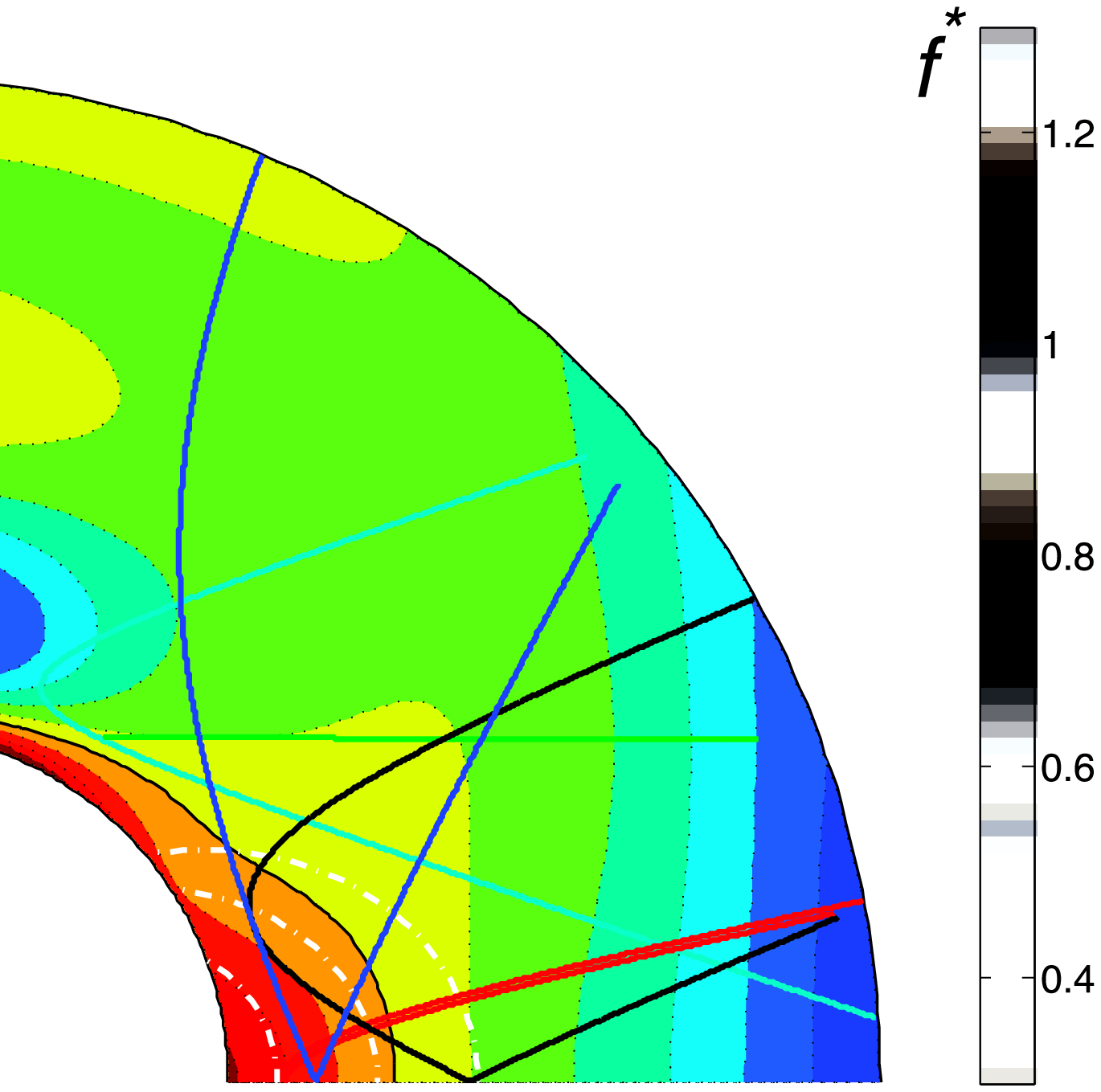}}
{\includegraphics[width=0.55\linewidth]{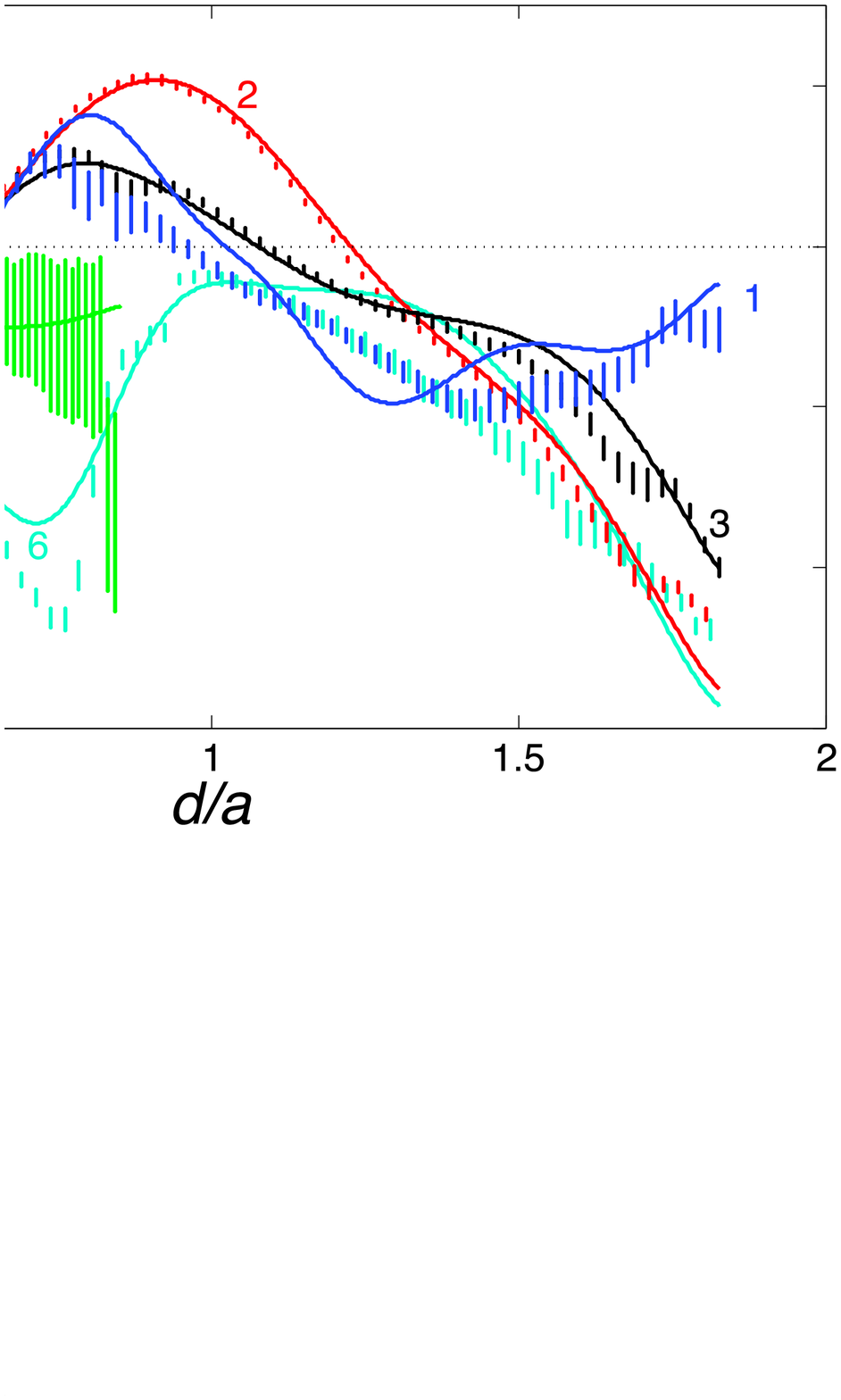}}}
\caption{(Color online) (a) Reconstructed isovalue map of fluid angular frequency $f^{*}$ (the fluid angular frequency normalized by $f$) at $f=\pm3$ Hz in a meridional plane, assuming axisymmetry and symmetry with respect to the equator. Three dipolar field lines (dash-dot white) are superimposed in the angular velocity maps. Super-rotation ($f^{*}$ > 1) is clearly visible near the inner sphere, where the Ferraro law
of isorotation applies. Contours become vertical further away, where geostrophy dominates. The fluid frequency
is higher than 0.4 everywhere except in thin unresolved boundary layers.
The color lines are the projection in the upper half $(s,z)$ plane of the ultrasonic rays used in the inversion (see FIG.~\ref{trajectoires}). 
(b) Comparison between the measured ultrasonic Doppler $f^{*}$ (shown by their error bars) and the synthetic profiles (solid lines)
computed from the angular frequency map of (a) for $f=\pm3$ Hz.
The $x$-axis gives the distance along the ray (in $a$ units).
The corresponding rays are plotted in (a) with the same color code (and indicated with trajectory numbers referring to FIG.~\ref{trajectoires}).}
\label{isolines}
\end{figure}

FIG.~\ref{isolines} shows the isovalues of angular frequency $f^{*}$  inverted for $f=\pm3$ Hz, with $l_{max}=3$ and $k_{max}=7$.
A crescent of super-rotation
is present near the inner sphere. There, isorotation contours roughly follow magnetic field lines, in agreement with Ferraro's
theorem, as anticipated above. At larger cylindrical distance from the inner sphere, the flow becomes geostrophic: the
contour lines are vertical. 
We note that angular velocities just above the north pole of the inner sphere do not comply with Ferraro's law.
Instead, velocities decrease to quite low values inside the cylinder tangent to the inner sphere.
Such violations have been shown to occur when the electric conductivity of boundaries is high \cite{allen1976law,soward2010shear}.
We speculate that we might be in this situation inside the tangent cylinder because the opening of the sphere at the top
and bottom (see FIG.~\ref{trajectoires}) replaces the poorly conducting stainless steel wall by sodium.

FIG.~\ref{isolines} compares the synthetic angular velocity profiles to the observed Doppler velocity profiles along the
various rays. Note that super-rotation is clearly visible in the raw profiles.
The drop in velocity just above the inner sphere is constrained by profiles 4 (green) and 6 (cyan), but its vertical extent is not.

\subsection{$f_{fluid}$ deduced from differences in electric potential and from UDV}

\begin{figure}[h]
\centerline{{\includegraphics[width=1\linewidth]{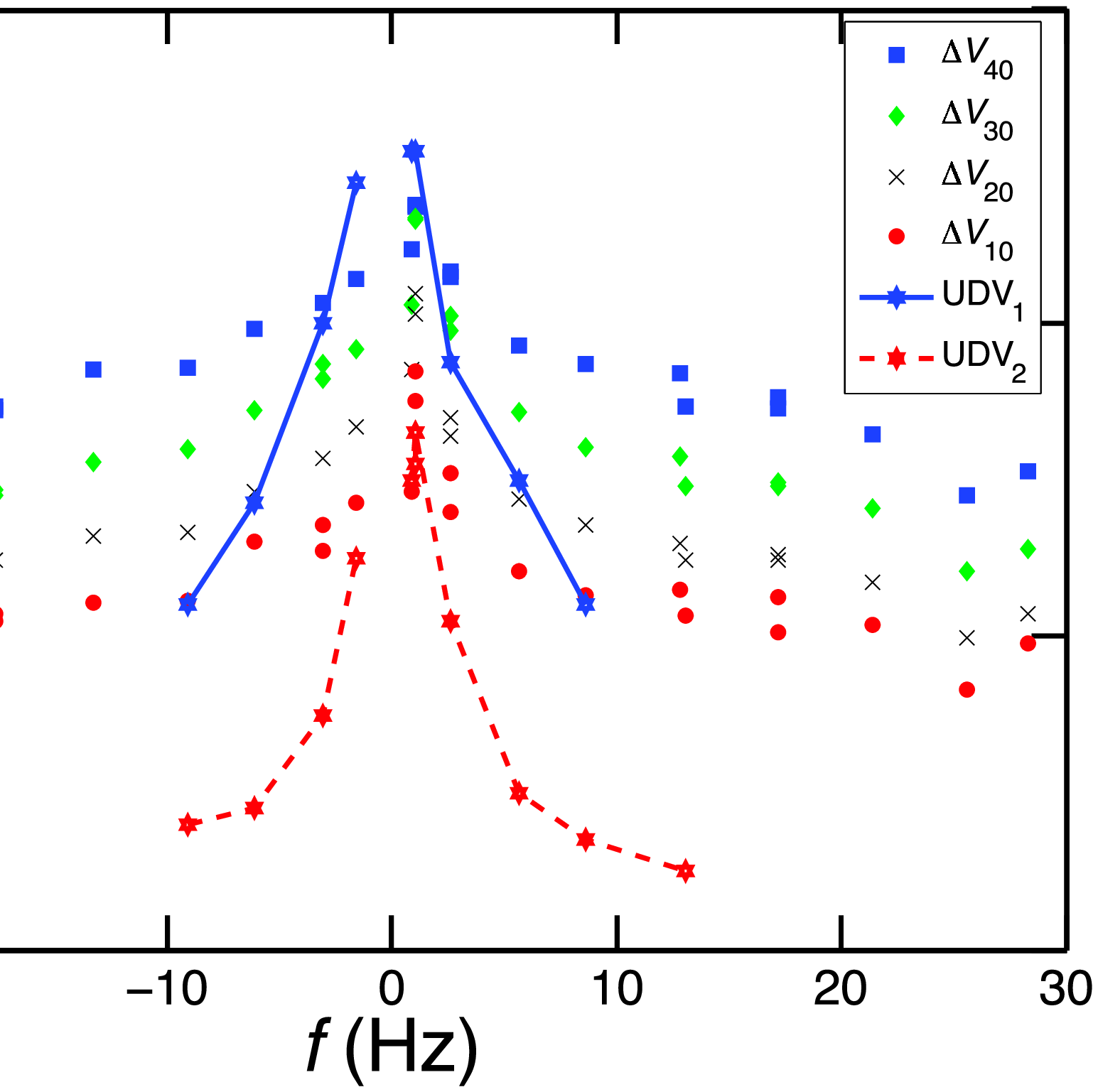}}}
   \caption{(Color online)  $ f^*$ deduced from the measurements
of $\Delta V$  using formula (\ref{Ohm2}). Blue  line : $ f^*$ value obtained with UDV measurements on the trajectory number 1 at the distance $d/a=0.1$. Dashed red line : $ f^*$ value obtained with UDV measurements on the trajectory number 2 for $d/a=0.1$. }
   \label{ddp}
\end{figure}

As in the previous study of $DTS$ with rotating outer sphere \cite{nataf07}, we observe that the amplitudes of the
differences in electric potential
$\Delta V$'s vary linearly with $\Delta V_{40}$, the proportionality factor increasing from the equator toward the poles due in particular to the increase of $B_{r}$ in formula (\ref{Ohm2}).
We show however in the present study that measuring the electric potential does not yield a reliable indicator of the angular velocity $f^{*}$ using formula (\ref{Ohm2}). 
In FIG.~\ref{ddp}, we compare the normalized fluid angular velocity $f^{*}$ retrieved from the $\Delta V$'s,  for four different latitudes, to $f^{*}$ obtained directly by UDV at the nearest measured point, around $d/a =  0.1$.  The frequencies $f^{*}$ obtained from $\Delta V$ and from UDV in FIG.~\ref{ddp}, would be similar if both measurement techniques were only sensitive to $u_{\varphi}$ in the interior below the outer viscous boundary layer.
The strong discrepancy between these two sets of frequencies reveals instead that the outer boundary layer in $DTS$ cannot simply be reduced to a Hartmann layer, outside of which the meridional currents $j_{\theta}$ can be neglected. We further discuss this point in the numerical part \ref{comparison}.  
 
\section{Meridional circulation}

The meridional circulation is constrained from Doppler velocity profiles of the radial velocity (shot along the radial
direction), from profiles shot in a meridional plane, and from the projection of the meridional velocity on "azimuthal"
shots. The latter is obtained by taking the sum of the profiles acquired for $f$ and $-f$, in order to eliminate
the azimuthal contribution. The same is done for the radial and meridional profiles to remove any contamination
from azimuthal velocities.

The poloidal velocity scalar $u_P$ of equation (\ref{pol-tor}) is expanded in associated Legendre functions with even degree and order 1, {\it i.e.}
\begin{equation}
u_P(r,\theta)=\sum_{l=0}^{l_{max}}u_P^l(r) P_{2l}^1(\cos\theta) \; .
\label{even-leg}
\end{equation}

The radial $u_r$ and orthoradial $u_{\theta}$ components of velocity are then obtained as:
\begin{equation}
u_r(r,\theta) = \sum_{l=0}^{l_{max}} \frac{u_P^l(r)}{r} \frac{1}{\sin \theta} \frac{d}{d \theta} \left( \sin \theta \; P_{2l}^1 (\cos \theta)\right).
\label{eq:u_r}
\end{equation}
\begin{equation}
u_{\theta}(r,\theta) = - \sum_{l=0}^{l_{max}} \left( \frac{u_P^l(r)}{r} + \frac{d u_P^l(r)}{d r} \right) \; P_{2l}^1 (\cos \theta).
\label{eq:u_theta}
\end{equation}

\begin{figure}[p]
\centerline{\includegraphics[clip=true,width=0.5\linewidth]{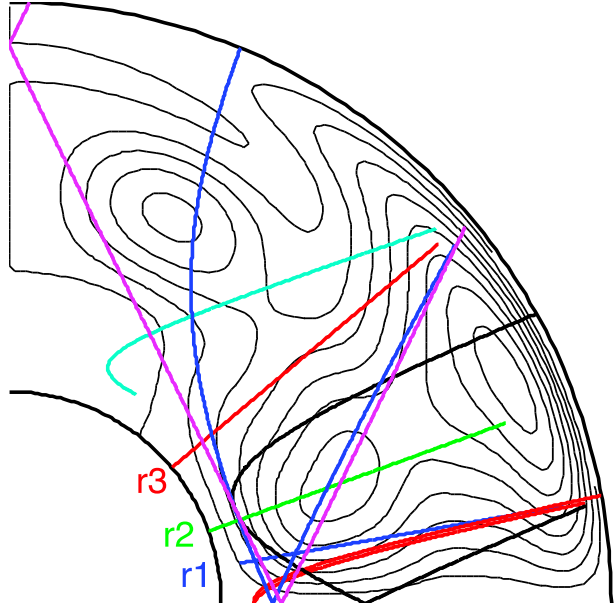}}
\caption{(Color online) Reconstructed stream lines of the meridional circulation at $f=\pm3$ Hz in a meridional plane, assuming axisymmetry and symmetry with respect to the equator. The interval between lines is $1.6 \times 10^{-3}$.
The fluid is centrifuged away from the inner sphere
in the equatorial region and moves up to the pole along the outer boundary.
The color lines are the projection in the upper half $(s,z)$ plane of the ultrasonic rays used in the inversion.}
\label{HCN_mer_stream}
\end{figure}

The functions $u_P^l(r)$ are decomposed into a sum of $\sin \left( k \pi (r-b/a)/(1-b/a) \right)$ from $k=0$ to $k_{max}$.
The radial velocity is thus constrained to vanish at the inner and outer (rigid) boundaries, but the orthoradial velocity is
not, in order to account for the presence of thin unresolved boundary layers. FIG.~\ref{HCN_mer_stream} shows
the streamlines of the meridional circulation inverted for $f = \pm 3$ Hz, with $l_{max} = 4$ and $k_{max} = 8$. The
fluid is centrifuged from the inner sphere in the equatorial plane and moves north in a narrow sheet beneath the outer boundary. It loops
back to the inner sphere in a more diffuse manner. Meridional velocities are more than ten times weaker than
azimuthal velocities.

\begin{figure}[h]
\centerline{{\includegraphics[width=0.88\linewidth]{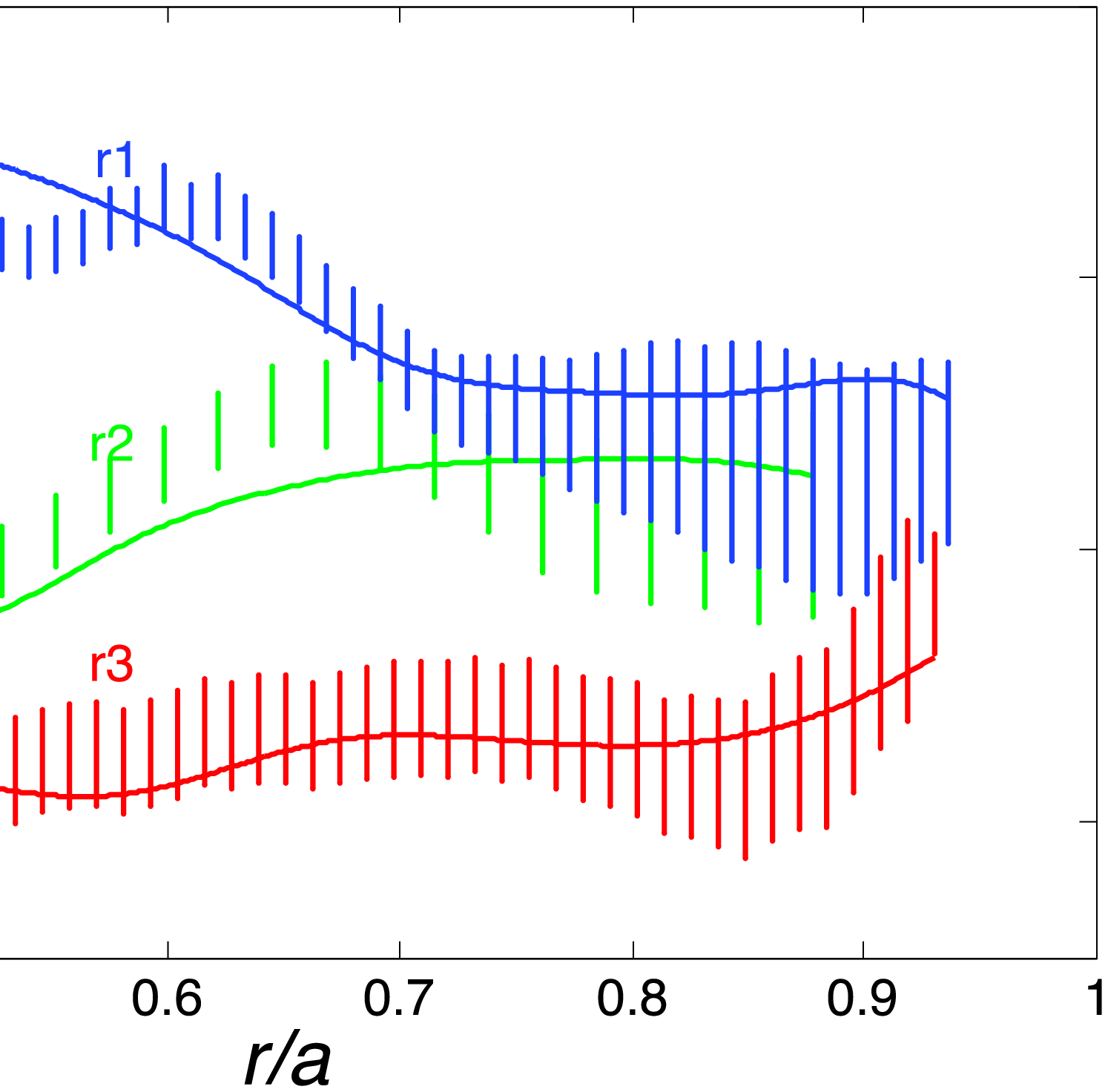}}}
\centerline{\hspace*{-0.25cm}{\includegraphics[width=0.915\linewidth]{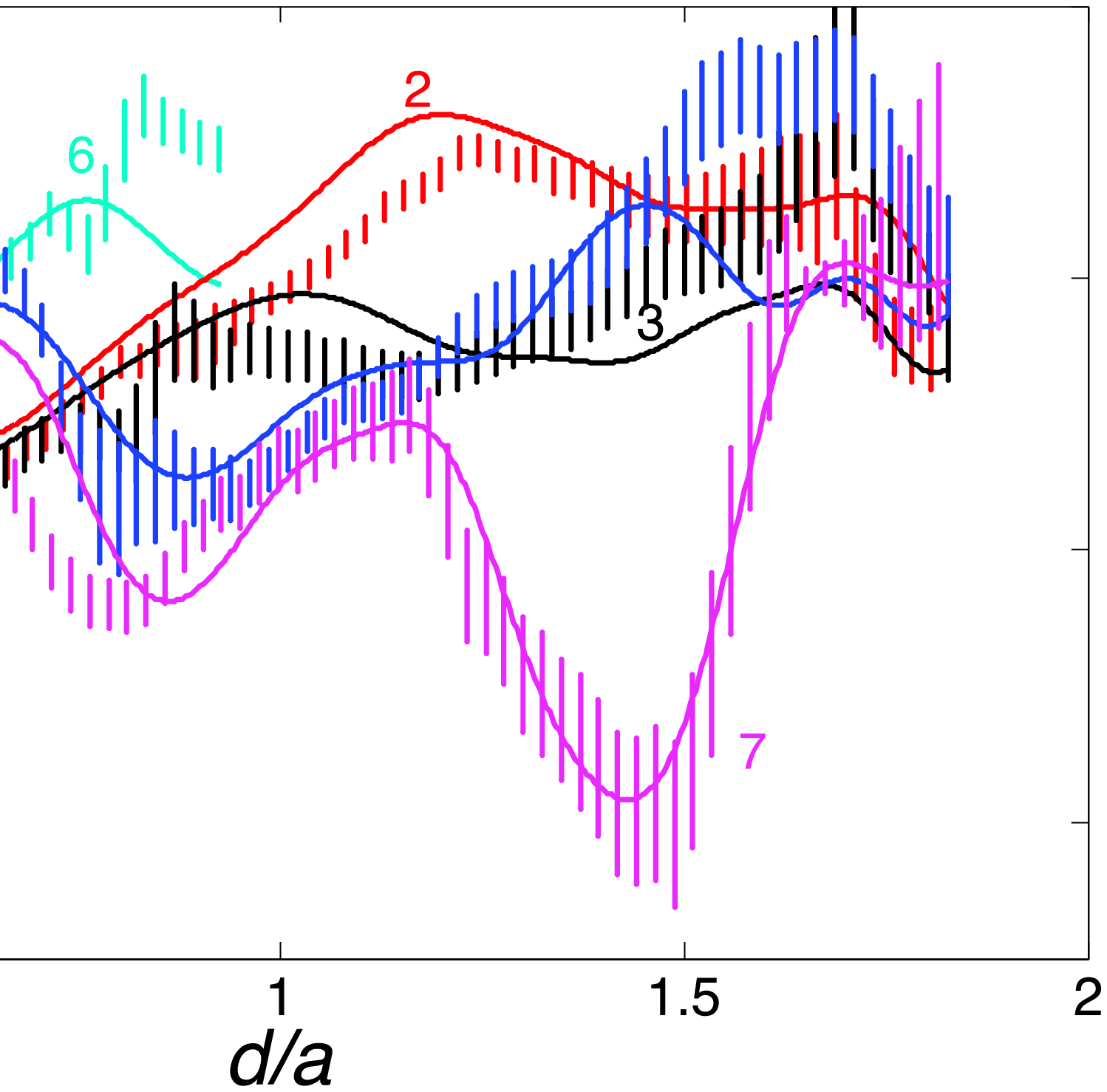}}}
\caption{(Color online) Comparison between the measured ultrasonic Doppler velocity profiles (shown by their error bars) and the synthetic profiles (solid lines)
computed from the meridional circulation map of FIG.~\ref{HCN_mer_stream} for $f=\pm3$ Hz. (a) Radial profiles along the radial direction r1, r2 and r3 shown in FIG.~\ref{HCN_mer_stream}. (b) "Azimuthal" profiles. The contribution 
from the azimuthal flow has been removed by taking the sum of profiles acquired for
$f$ and $-f$. The $x$-axis gives the distance along the ray (in $a$ units) and the $y$-axis is the velocity measured along the ray, adimensionalized by $2 \pi f a$. The corresponding rays are plotted in FIG.~\ref{HCN_mer_stream} with the same color code (for the grayscale version, the trajectory numbers in (b) refers to those in FIG.~\ref{trajectoires}).}
\label{HCN_mer_prof}
\end{figure}

FIG.~\ref{HCN_mer_prof} compares the synthetic radial and meridional profiles to the observed Doppler
velocity profiles along the various rays. Velocities are normalized by $2\pi f a$.

\begin{figure}[h]
\centerline{{\includegraphics[width=0.88\linewidth]{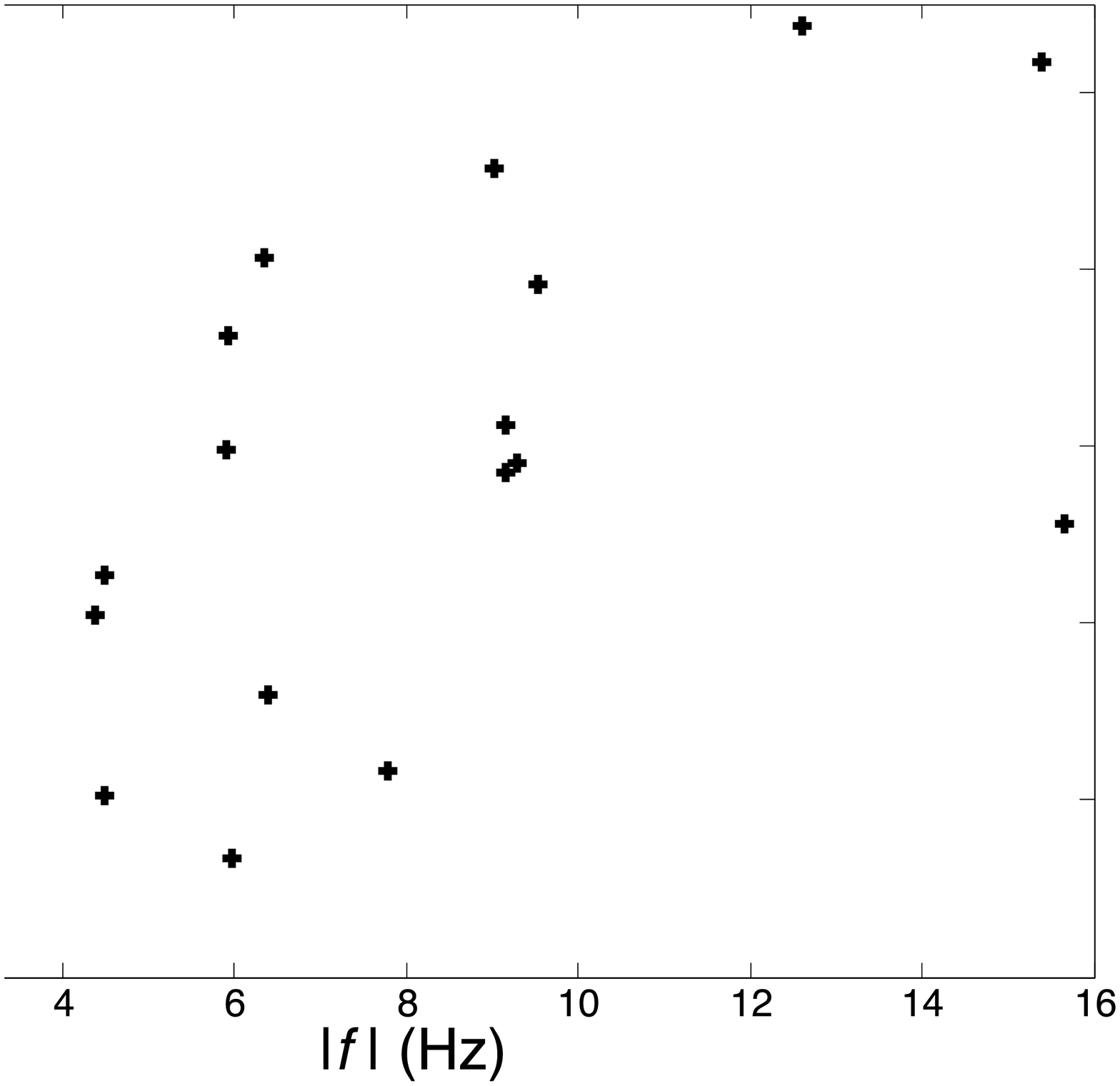}}}
\caption{Compilation of the $rms$ radial velocity amplitude as a function of the absolute value of $f$.
The $rms$ velocity is computed from Doppler velocimetry profiles shot at a latitude of $20\deg$, from 3 cm beneath the outer shell down to the inner sphere (to avoid spurious values close to the outer boundary). 
Radial velocity is roughly proportional to $f$ but there is a large dispersion, as the shape of the profiles changes with $f$.
Note that for $f=10$ Hz, the tangential velocity on the inner sphere reaches 465 cm/s.}
\label{HCN_Ur_rms}
\end{figure}

Over a decade (from $f=1.5$ Hz to $=15$ Hz), radial velocities are consistently centrifugal at $10\deg$ latitude and
centripetal at $40\deg$, and are roughly proportional to $f$. The radial profiles at $20\deg$ are more complex
and evolve with $f$, indicating a non-monotonic evolution of the meridional circulation, also evidenced by the
records of the $r$ and $\theta$ components of the induced magnetic field inside the fluid (see FIG.~\ref{inducedB}).
FIG.~\ref{HCN_Ur_rms} compiles the $rms$ value of radial velocity at $20\deg$ for various $f$.
Note that the fluctuations are larger than this value, which is almost 50 times smaller than azimuthal velocities.

\section{Comparison with numerical simulations}
\label{comparison}

Two previous numerical studies are particularly relevant to our work. Hollerbach et al. studied exactly the $DTS$ configuration but for values of $\Lambda$ much larger than its value in the experiment \cite{hollerbach07}. They focus their study on the modification of the linear solution by inertial effects, stressing that the magnetic field line tangent to the outer sphere at the equator loses its significance in the non linear regime. As a result of the relatively large value of $\Lambda$, the inertial effects remain too weak -when the outer sphere is at rest- to make a geostrophic region arise at large distances from the axis. The solutions of Garaud \cite{garaud2002} (see the figures 7 and 11) for a slightly different problem do show the transition between a Ferraro and a geostrophic regions. In her model, which pertains to the formation of the solar tachocline, a dipolar magnetic field permeates a thick spherical shell as in $DTS$, the rotation of the outer boundary is imposed and the rotation of the inner boundary is a free parameter: a condition of zero torque is imposed on that boundary. Numerical models \cite{hollerbach07,nataf07} of the $DTS$ experiment when the outer sphere is rotating also clearly show a Ferraro region near the inner sphere where the magnetic field is strong and a geostrophic region in the vicinity of the equator of the outer sphere. We argue below that all these results obtained for a rotating outer sphere provide us with a useful guide to interpret the numerical solutions when the outer sphere is at rest.

\subsection{The numerical model}
\begin{figure}[htbp]
\begin{center}
\includegraphics[clip=true,width=0.5\linewidth]{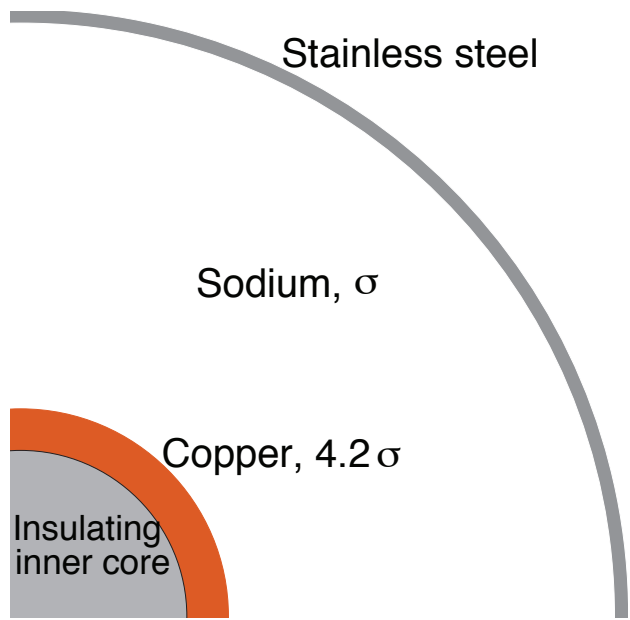}
\caption{(Color online) Geometry of the numerical model. The relative conductance of the solid outer shell is  $\sigma_b\delta/\sigma a=1/336$, with $\sigma_b$   and $\delta$ respectively the conductivity and the thickness of the outer sphere. It reproduces the experimental value with $\sigma_b$ chosen as the conductivity of stainless steel at 140$\deg$C. The conductivity ratio between the layers 2 and 3 reproduces the ratio (4.2) between the conductivity of copper and sodium.}
\label{model_num}
\end{center}
\end{figure}

The model consists of four nested spherical layers (see FIG.~\ref{model_num}). The fluid layer is enclosed between a weakly conducting outer container and a central solid sphere comprised of an inner insulating core and of a strongly conducting outer envelope.

The velocity field is decomposed as stated in the definitions (\ref{pol-tor}) and (\ref{odd-leg}).
The variables $u_\varphi^l(r)$ and $u_p^l(r)$ are then discretized in radius.
Analogous decompositions of variables denoted $b_\varphi^l(r)$ and $b_p^l(r)$ are employed to represent the induced magnetic field.
The truncation level $l_{max}$ (see (\ref{odd-leg})) is 120 and at least 450 unevenly spaced points are used in the radial direction.
Specifically, the density of points strongly increases close to the boundaries in order to resolve the viscous boundary layers.

The equations (\ref{motion}) and (\ref{induction}), modified to include all the non linearities and the time derivatives of $\mathbf{u}$ and $\mathbf{b}$, are transformed into equations for $u_\varphi^l$, $u_p^l$, $b_\varphi^l$ and $b_p^l$. 
We treat the non linear terms explicitly. To advance from one time step to the next, we use an Adams-Bashforth method. Diffusive terms, however, are treated implicitly.
Finally, Laplace's equation in spherical coordinates separates which makes it easy to write the magnetic boundary conditions.

The dimensionless numbers $\Re$ and $\Lambda$ are chosen so that steady solutions exist and are stable, with $\Pm\ll 1$ ($\Pm$ enters the definition of the unit induced field). We strive to reproduce the experimental values of $\Lambda$ and $R_m$.
Solutions are obtained after time-stepping the equations until a stationary or periodic state is reached. They have been successfully compared to solutions obtained with another numerical code PARODY, which is not restricted to axisymmetric variables \cite{aubert2008,guervilly10}.

It is not possible to simulate the Reynolds number of the experiment, which is about $10^6$. For the experimental range of $\Lambda$, steady solutions are obtained with $\Re\sim 10^3$.

\subsection{Steady axisymmetric solutions}

\begin{figure}[h]
\centerline{\includegraphics[clip=true,width=0.36\linewidth]{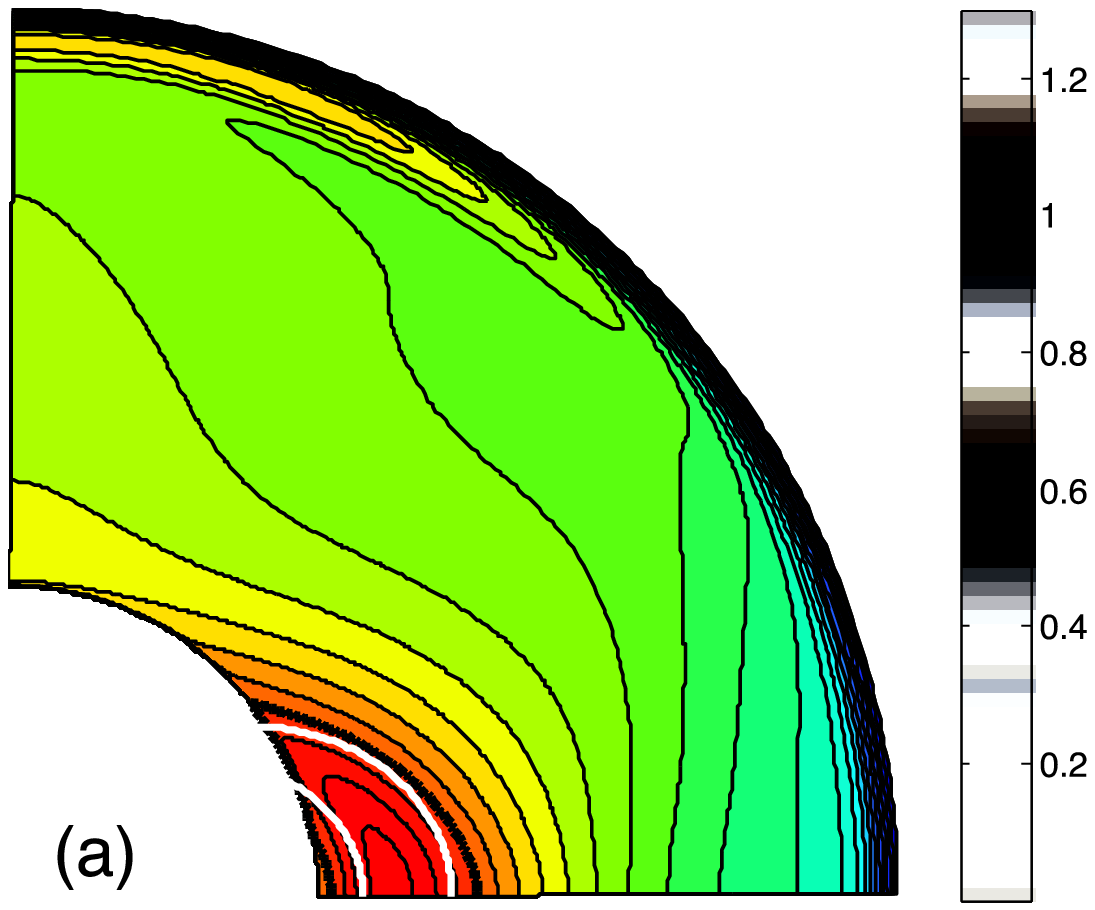}
\includegraphics[clip=true,width=0.30\linewidth]{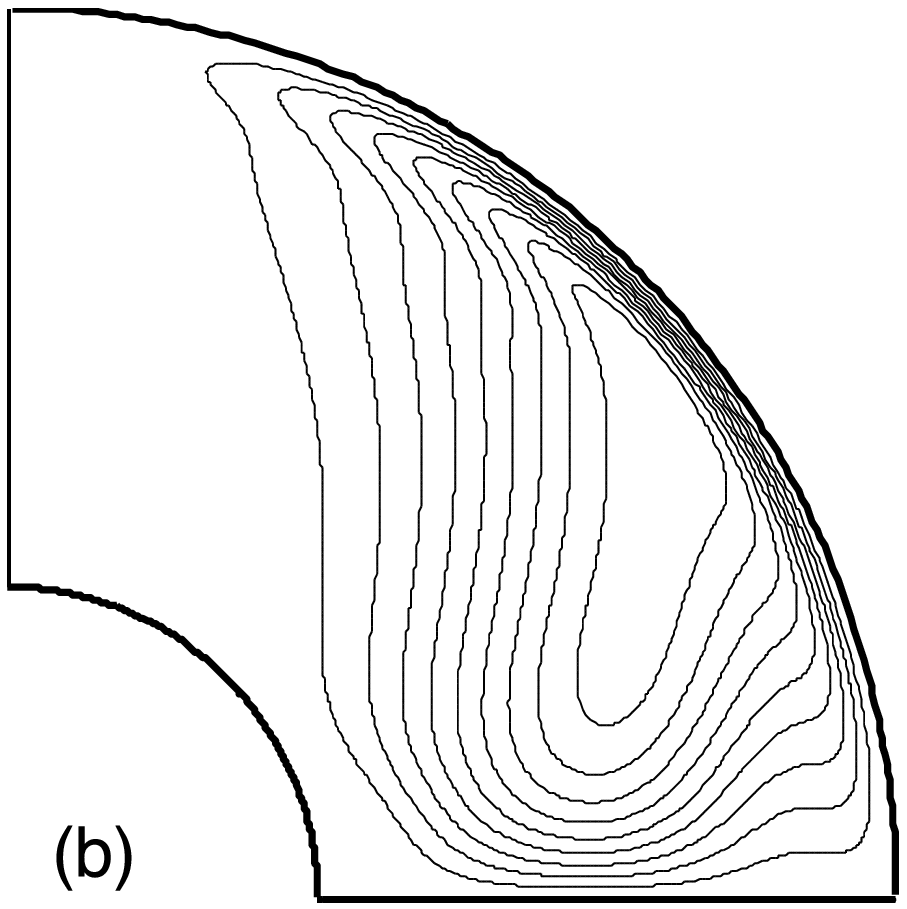}
\includegraphics[clip=true,width=0.31\linewidth]{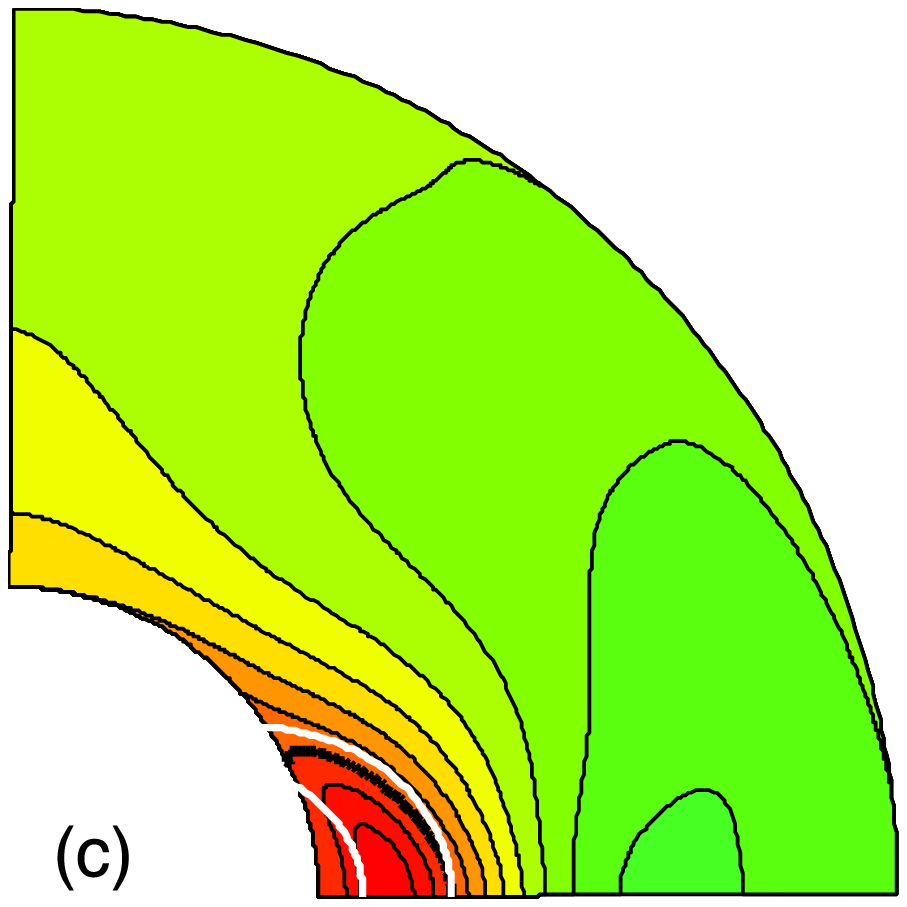}}
\caption{(Color online) (a) Angular and (b) meridional  velocity in a meridional plane for $\Re=9.5\;10^2$, $\Ha=163$, and $\Pm=10^{-3}$.
(c) angular velocity estimated from $V$, using (\ref{Ohm2}).
Two dipolar field lines (white) are superimposed in the angular velocity maps, and the thick black contour line is where
the angular velocity is unity.}
\label{solref}
\end{figure}

FIG.~\ref{solref} displays a typical solution for the angular and meridional velocities that illustrates well the experimental results. The fluid rotates faster than the magnetized inner body in its vicinity. There, the angular velocity is constant along magnetic field lines of force. Further away of the inner core, the zonal shear becomes almost geostrophic. In addition to these features that we have retrieved from the experimental results, the numerical solution displays recirculation in the outer boundary layer at high latitude. There, the interior flow largely consists in rigid rotation and the boundary layer has the characteristics of a B\"odewadt layer with a region of enhanced angular rotation.

For large enough $\Re$ ({\it e.g.} $(a/b)^2 \Re = 10^4$ with $(b/a)^2 \Ha = 20$), circular waves are present in the  B\"odewadt layer, above $60^{\circ}$ of latitude. They propagate towards the axis. Similar waves had been reported before in simulations of the flow between a rotating and a stationary disk in the absence of a magnetic field \cite{lopez2009crossflow}. There, they eventually die out. Thus, the persistence of propagation of circular waves in the boundary layer attached to the sphere at rest may be attributed to the presence of a magnetic field. On the other hand, these waves arise for larger $\Re$ as $\Ha$ is augmented. Their emergence delimits the domain of steady solutions.

We have checked that the thickness of the outer boundary layer in the numerical solution scales as $\Omega^{-1/2}$. Note that it corresponds to 3  mm for $\Omega=1.5$ s$^{-1}$ and the viscosity of liquid sodium. The fluid rotation is driven by the electromagnetic torque acting at the inner boundary against the viscous torque at the outer boundary. We have found that both the viscous torque on the inner surface and the electromagnetic torque on the outer surface are negligible. Comparing different simulations, we have also checked that the main viscous torque scales as $\sim\Omega^{3/2}$, as expected from the thickness of the B\"odewadt layer. Thus, the power required to drive the fluid rotation scales as $\Omega^{5/2}$, as does the experimentally measured power, and torque measurements do not give indications on turbulence (see section \ref{Powsca}).

The angular rotation just below the outer viscous layer scaled by the inner core angular rotation decreases with $\Re$ in agreement with the experimental results. On the other hand, the angular rotation that would be inferred from the electric potential differences calculated at the outer surface using expression (\ref{Ohm2}) increases with $\Re$.
FIG.~\ref{solref}(c) displays the angular velocity as estimated from the electric potential, according to equation (\ref{Ohm2}). It can be compared to FIG.~\ref{solref}(a). The actual shear is well retrieved where the magnetic force predominates, in the
region where Ferraro's law of isorotation holds. There, the electric current density $\mathbf{j}$ is limited by the strength of the magnetic force, which needs to be balanced by another force. That restriction makes it possible to neglect $\mathbf{j}$ in Ohm's law. Then, predictions made from (\ref{Ohm2}) are correct. On the other hand, the actual shear is not well recovered in the geostrophic region where the electric current density is not limited by the strength of the magnetic field.
There, the frozen-flux relation (\ref{Ohm2}) can be violated. We thus explain why the electric potential measurements at the surface of the $DTS$ experiment do not yield a good prediction of the angular velocity immediately below the outer viscous boundary layer.

Our first discussion \cite{nataf06} of the electric potential measurements was based on a numerical model calculated for the experimental values
of $\Ha$ and thus for too large values of $\Lambda$. As a result, the magnetic force, in the numerical model, was dominant in the entire fluid layer and
the frozen-flux relationship (\ref{Ohm2}) was verified, at least away from the equator where $B_r=0$.  However, equation (\ref{Ohm2}), becomes less
and less valid as $\Re$ is increased and $\Lambda$ decreased, in agreement with the divergence that has been experimentally observed (see the FIG.~\ref{ddp}) between the angular velocity calculated from (\ref{Ohm2}) and the actual velocity.

Incidently, cranking up the rotation of the magnetized inner sphere stabilizes the fluid circulation, at least within a certain parameter range. We have calculated the time-averaged solution (not shown) for the same parameters as the steady solution illustrated by FIG.~\ref{solref}, but for a lower $\Re$. Both the flow and the induced magnetic field are periodic for this set of parameters. A second meridional roll, which is centripetal in the equatorial plane, turns up in the outer region. There, it creates a disk-shaped region where the rotation is slow and the solution is strikingly different from the almost geostrophic solution (FIG.~\ref{solref}) obtained for a slightly larger value of $\Re$.

\clearpage
\subsection{Comparison between numerical simulations and experimental results}

We find that reproducing the Elsasser number $\Lambda$, rather than a combination of $\Lambda$ and $\Re$ such as the Hartmann number $\Ha=(\Re\Lambda)^{1/2}$, is the key factor to recover the experimental results. The parameters for the solution displayed in FIG.~\ref{solref} correspond to $\Lambda = 28$, which is the appropriate value for experiments with $\Omega=1.5$ s$^{-1}$. With $\Pm=10^{-3}$, the value of the magnetic Reynolds number is about right. It remains too small for the poloidal field to be much different from the imposed dipole field (again for the parameters of FIG.~\ref{solref}).

\begin{figure}[h]
\begin{center}
\includegraphics[clip=true,width=1\linewidth]{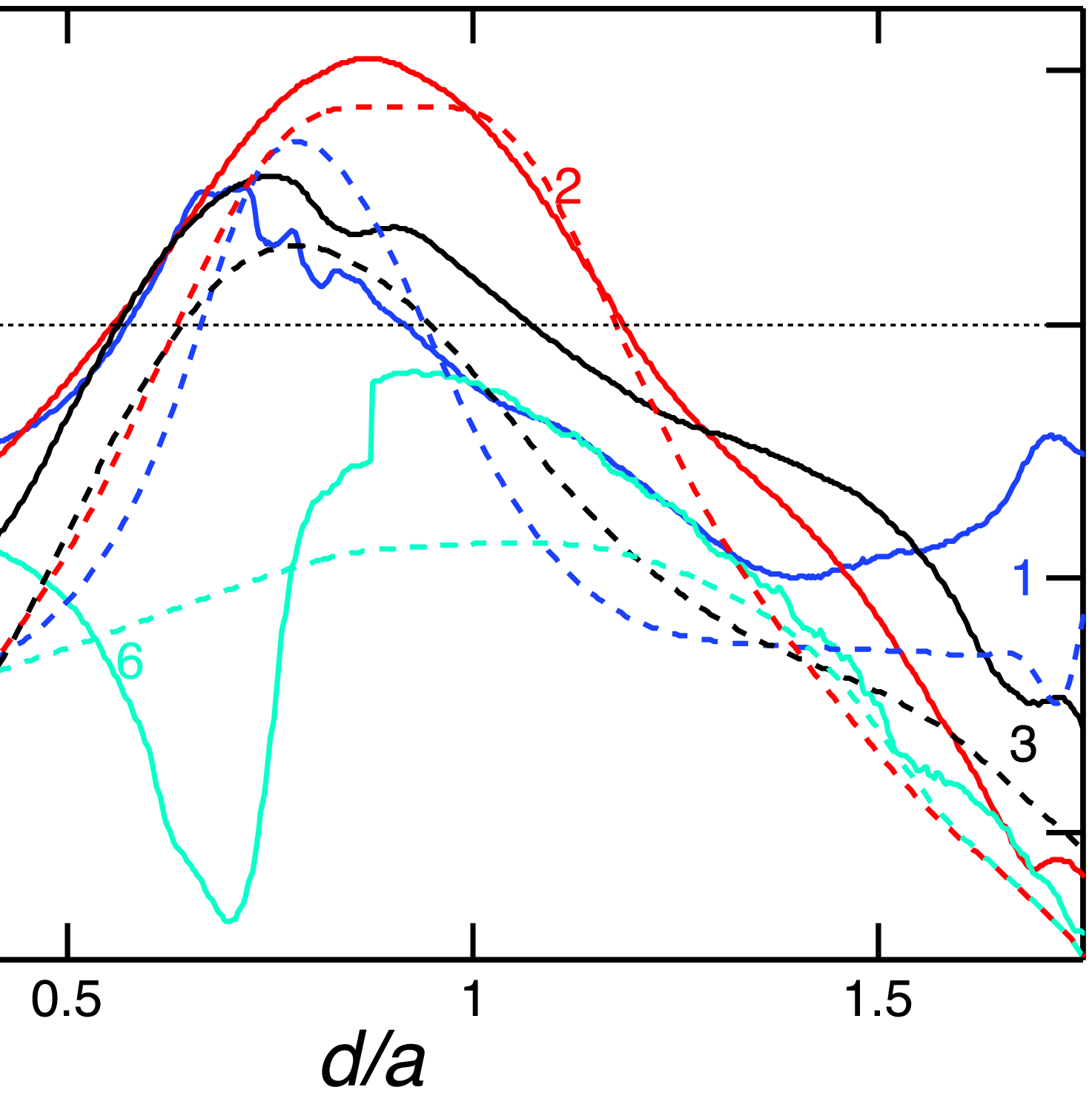}
\caption{(Color online) Angular velocity along the ultrasonic rays as a function of the distance from the probe: measured (solid lines, 3 Hz, $\Lambda=16$) and retrieved from a time-averaged numerical solution (dashed lines,
$\Re=1.5\;10^3$, $\Ha=163$, $\Pm=10^{-3}$, $\Lambda=18$). The color lines refers to those used to define the ultrasonic beams in the FIG.~\ref{trajectoires} (the numbers also refers to the ultrasonic beams numbers defined in the FIG.~\ref{trajectoires}).
The error bars of the experimental data are shown in FIG.~\ref{isolines}.}
\label{comp_dop}
\end{center}
\end{figure}

FIG.~\ref{comp_dop} shows that numerical solutions are able to satisfactorily reproduce the ultrasonic measurements of angular velocity, obtained for the same values of $\Lambda$, as expected from the similitude of the angular velocity maps \ref{isolines} and \ref{solref}. The simulated velocities have weaker amplitude than the measured ones in much of the fluid though. We have checked that increasing $\Re$, whilst keeping $\Lambda$ constant, favours enhanced corotation between the fluid and the inner core. As our calculations are for much smaller $\Re$ than the values realized in the experiment, that result may explain the remaining discrepancy between measured and simulated velocities.

\clearpage
\section{Discussion and conclusion}
\label{conclusion}
In the presence of an imposed magnetic field, which favors solid body rotation, the inertial forces largely reduce to a Coriolis force, even for large Reynolds numbers. Experimental results can thus be interpreted using a single dimensionless number, the Elsasser number. In that respect, experimental results obtained with global rotation \cite{nataf07} provide a better guide
to interpreting the present results than the linear situation studied by Dormy et al. \cite{dormy98,dormy02}. 
We estimate that, in $DTS$, the rotation frequency $f$ should be less than $0.1$ Hz for the latter to be approached.

Experiments have been conducted with the inner sphere rotating in the range -30 Hz $\le f\le$ 30 Hz.
We have been able to map extensively the shear in the fluid cavity from ultrasonic Doppler velocimetry for $|f| \le$ 10 Hz.
Our observations provide a very clear experimental illustration of Ferraro's law of isorotation, demonstrating the predominance of magnetic forces near the inner sphere.
They also exhibit a strong super-rotation: in the region where magnetic forces dominate, the fluid angular velocity gets 30$\%$ larger than that of the inner sphere.
This contrasts with the results obtained by Dormy et al. \cite{dormy98} when global rotation is present, which indicate that the phenomenon of super-rotation is hindered by the Coriolis force.
The experimental results obtained in our previous study with global rotation \cite{nataf07} could not address this issue and we plan to run additional experiments for that purpose.

The experiments also display a clear violation to Ferraro's law: quite low angular velocities are observed just above the inner sphere, where the magnetic field is strongest (see FIG.~\ref{isolines}).
We suspect that this is due to the presence of sodium at rest at the top and bottom of the cylinder tangent to the inner sphere. 
Indeed, such violations have been shown to occur when the electric conductivity of boundaries is high \cite{allen1976law,soward2010shear}.

We could follow the evolution of induced magnetic field, electric potentials and power across the full range of forcing.
In a first approximation, all observables associated with the azimuthal flow (which dominates) can be described by a universal solution, both velocities and induced magnetic field scaling with $f$. 
In a second approximation, the increase of the dimensional fluid velocity with $f$ thins the viscous boundary layer at the outer sphere and increases friction accordingly, thus reducing the adimensional velocity of the fluid inside the sphere.
At the same time, the effective Coriolis force that results from the non-linear $(\mathbf{u}\cdot \mathbf{\nabla})\mathbf{u}$ term increases with respect to the (linear) Lorentz force: the geostrophic region extends further towards the inner sphere.
This explains that the fluid velocity increases with $f$ less rapidly than $f$ (FIG.~\ref{VitessePaper}) at large $f$ whilst the torque instead increases more rapidly than $f$ (FIG.~\ref{ElectrodesPaper}) (the electric potentials follow an intermediate trend).
The outer friction torque is balanced by the magnetic torque at the inner boundary.
This is consistent with an increase of the induced magnetic field, near the solid inner body, that is steeper than $f$ (see FIG.~\ref{inducedB}).
On the other hand, the description of Nataf and Gagni\`ere \cite{Nataf:2008fk} pertains to the region where the shear is geostrophic.
There, the increased torque at the outer boundary is balanced by the magnetic torque on the geostrophic cylinders in the interior, which results from the shearing of the imposed dipolar field.
The direct measurement of the velocity (up to 10 Hz, see FIG.~\ref{geostrophy}) shows that the adimensionalized shear does not change significantly with $f$ even though the velocity itself decreases. 
In addition, the induced azimuthal magnetic field that we measure inside the sphere (FIG.~\ref{inducedB}), for the whole range of $f$,
increases more rapidly than $f$.
At large $f$, we observe that $b_{\varphi}$ gets larger than the imposed dipolar field in much of the fluid layer.
Eventually, this induced field is large enough to modify the overall magnetic field, and the resulting flow.

This last regime, only achieved because the magnetic Reynolds number is large enough, is probably the most interesting one.
Unfortunately, we cannot directly measure the flow velocities with the ultrasound technique at these very large $f$. 
Less direct techniques are now required to investigate the zonal shear for $f >$ 10 Hz. 
Inertial waves modified in the presence of the dipolar and the induced magnetic fields have been inferred from records of the electric potential along parallels at the surface \cite{schmitt07} and of the magnetic field 
along a meridian. 
Both their period and their wavenumber vary with the geometry of the differential rotation in the cavity. 
Hopefully, it will be possible to invert the zonal shear from the records of magneto-inertial waves.

Guided by the numerical model, we find that electric field measurements are difficult to interpret, particularly in the equatorial region where the radial magnetic field $B_r$ vanishes. 
The frozen-flux approximation (\ref{Ohm2}) holds when there is a mechanism that keeps under control the strength of the electrical currents \cite{jackson1975classical}. This is the reason why the magnetic Reynolds number $\Rm$ is not relevant to discuss the validity of the frozen-flux approximation in our quasi-steady experiment. That approximation has predictive power, instead, in regions where the magnetic force is dominating. In the $DTS$ experiment, it corresponds to the inner region close to the magnet where $\Lambda \ge 1$.

In a geophysical context, a similar approach is routinely used \cite{TOG8holme07} to invert the velocity field at the Earth's core surface from models of the time changes of the geomagnetic field, the so-called secular variation. 
Taking the example of a quasi steady state, this geophysical application has been criticized from a strictly kinematic standpoint \cite{love1999critique}.
We reckon instead that it is necessary to consider the balance of forces to decide whether the frozen-flux hypothesis holds, at least for a quasi steady state as illustrated by the $DTS$ experiment.

Features of the experiment that only depend upon dimensionless numbers that do not involve diffusivities have been simulated numerically. An analogous explanation has been put forward to explain the intriguing successes of geodynamo simulations \cite{christensen2006}.

\begin{acknowledgments}
The $DTS$ project has been supported by Fonds National de la Science, Agence Nationale de la Recherche (Research program VS-QG, grant number BLAN06-2.155316), Institut National des Sciences de l'Univers, Centre National de la Recherche Scientifique, and Universit\'e Joseph-Fourier. We are thankful to Dominique Grand and his colleagues from $SERAS$ who conducted the design study of the mechanical set-up. The magnetic coupler was computed by Christian Chillet. We thank two anonymous referees for useful comments.
\end{acknowledgments}

\clearpage

\appendix*
\label{Appendix}
\section{Angular and meridional velocity along the ultrasonic oblique rays}

The seven oblique ultrasonic rays shot in $DTS$ are sketched in FIG.~\ref{trajectoires}. We define the declination $D$ as the angle between the  beam and the meridional plane ($D$ counted positively eastwards), the inclination $I$ as the angle between the projected beam in the meridional plane and the radial direction ($I$ counted positively upwards) and $\lambda$ as the latitude of the ultrasonic probe. Using those definitions, TABLE~\ref{trajectories} give the characteristics of the beams. 

\begin{table} [h]	
\caption{\label{trajectories} Latitude $\lambda$, inclination $I$, and declination $D$ (in degrees) at the origin of the
shots (on the outer sphere) of the oblique ultrasonic beams in $DTS$.}
\begin{ruledtabular}
\begin{tabular}{cllcc}
 Trajectory number and color  & $\lambda$  & $I$& $D$  \\
 \hline
{1, blue} & 40 & 21.1 & 11.7 \\
{2, red} & 10 & 2.2 & 23.9 \\
3, black& 10 & 12.5& -20.6 \\
{4, green} & -20 & 20& -13.5 \\
{5, yellow} & -20 & 21.1& -11.7 \\
{6, cyan} & -40 & 21.1& 11.7 \\
{7, magenta} & -40 & -24 & 0 
\end{tabular} 
\end{ruledtabular}
\end{table}

\subsection{Angular velocity}
Along these oblique beams, the projection $u(d)$ ($d$ is the distance from the probe) of the velocity  is a combination of the components $u_r$, $u_\theta$ and $u_\varphi$ of the total velocity field. 
Velocity $u(d)$ is counted positive in the shooting direction.
We assume that the mean fluid flow is
axisymmetric, and also ($u_r$,$u_\theta$) $\ll$
$u_\varphi$, the meridional velocities amplitude in $DTS$ being less
than  10$\%$ the amplitude of the azimuthal velocities. Using projections along the beam, we retrieve
the angular velocity $\omega(d)$  along trajectories 1 to 6
using the following relationship
\begin{equation}
\omega(d) = -\frac{u(d)}{a \cos \lambda \sin D}.
\label{delta_fluid}
\end{equation}
\noindent

\subsection{Meridional velocity}

We have also exploited the observation that the meridional velocity does not change sign when the rotation of the inner sphere is reversed - it remains centrifugal in the equatorial plane - whereas the angular velocity does change sign. Thus, combining measurements obtained with two opposite rotation rates of the inner core, we can separate azimuthal and meridional velocities.

Assuming now that the mean meridional velocity is axisymmetric and using projections, we can retrieve the radial velocity 
\begin{equation}
u_{r}(d) = \frac{u(d)r(d)}{d - a \cos  D \cos I}, 
\end{equation}
\noindent
and the orthoradial velocity
\begin{equation}
u_{\theta}(d) = \frac{u(d)r(d)s(d)}{a [ a \cos  D \cos \lambda \sin I -d \cos^{2} D \cos(\lambda+I) \sin I + d \sin ^{2} D \sin \lambda]}, 
\end{equation}
\noindent
where $r(d)=\sqrt{x^2+y^2+z^2}$ is the spherical radius and $s(d)=\sqrt{x^2+y^2}$ is the cylindrical radius at the measurement point.
They $(x, y, z)$ coordinates of the measurement point are given by:
\begin{eqnarray}
x(d) &=& a \cos \lambda - d \cos D \cos(\lambda +I) \\
y(d) &=& -d \sin D \\
z(d) &=& a \sin \lambda - d \cos D \sin(\lambda +I)
\end{eqnarray}

\bibliography{biblio}

\end{document}